\documentclass[reprint, nofootinbib]{revtex4-1}

\usepackage{amsmath}
\usepackage{amsfonts}
\usepackage{amssymb}

\usepackage{graphicx}

\usepackage[dvipsnames]{xcolor}
\usepackage{comment}

\usepackage[normalem]{ulem}

\newcommand{\pvalue}{\ensuremath{p\textrm{-value}}}
\newcommand{\pvalues}{{\pvalue}s}

\newcommand{\result}[1]{#1}

\begin{document}

\title{
A Coincidence Null-Test for Poisson-Distributed Events
}

\author{Reed Essick}
\email{reed.essick@gmail.com}
\affiliation{Kavli Institute for Cosmological Physics, The University of Chicago, 5640 South Ellis Avenue, Chicago, Illinois, 60637, USA}

\author{Geoffrey Mo}
\affiliation{LIGO Laboratory, Massachusetts Institute of Technology, 77 Massachusetts Avenue, Cambridge, Massachusetts, 02139, USA}
\affiliation{Department of Physics and Kavli Institute for Astrophysics and Space Research,
Massachusetts Institute of Technology, 77 Massachusetts Avenue, Cambridge, Massachusetts, 02139, USA}

\author{Erik Katsavounidis}
\affiliation{LIGO Laboratory, Massachusetts Institute of Technology, 77 Massachusetts Avenue, Cambridge, Massachusetts, 02139, USA}
\affiliation{Department of Physics and Kavli Institute for Astrophysics and Space Research,
Massachusetts Institute of Technology, 77 Massachusetts Avenue, Cambridge, Massachusetts, 02139, USA}

\begin{abstract}
When transient events are observed with multiple sensors, it is often necessary to establish the significance of coincident events.
We derive a universal null test for an arbitrary number of sensors motivated by the archetypal detection problem for independent Poisson-distributed events in gravitational-wave detectors such as LIGO and Virgo.
In these detectors, transient events may be witnessed by myriad channels that record interferometric signals and the surrounding physical environment.
We apply our null test to a broad set of simulated gravitational-wave events as well as to a real gravitational-wave detection to determine which auxiliary channels do and do not witness real gravitational waves, and therefore which are safe to use when constructing vetoes.
We also describe how our approach can be used to study detector artifacts and their origin, as well as to quantify the statistical independence of candidate GW signals from noise artifacts observed in auxiliary channels.
\end{abstract}

\maketitle

\section{Introduction}
\label{sec:introduction}

Establishing correlations between separate sets of possibly related transient events, or demonstrating the vanishingly small probability of observing a set of coincidences in uncorrelated events, is of general interest.
This situation arises naturally in a variety of contexts in astrophysics, especially in the correlation of events across multiple detectors.
Examples include the multiple elements of cosmic-ray arrays (e.g., \cite{PierreAuger, HAWC}); neutrino detections from IceCube~\cite{IceCube}, Antares~\cite{Antares}, and the Supernova Early-Warning System (SNEWS~\cite{SNEWS}); correlated magnetic noise from Schumann resonances in distant detectors \cite{Schumann1, Schumann2, Thrane2013}; as well as coincidences between gravitational waves (GWs), electromagnetic waves, cosmic rays and neutrinos in multimessenger astrophysics (see e.g.~\cite{GW170817MMA}).
Additionally, such questions are often asked while characterizing noise artifacts in detectors and searching for their sources (e.g.,~\cite{Biswas2013, Essick2013, Essick2020}).
This problem also appears outside the physical sciences, for example, in understanding internet server loads and distributed denial-of-service (DDOS) attacks (e.g.,~\cite{Bhandari2016}), aspects of quantitative finance (such as identifying causation in price impact~\cite{Bouchaud2009} or factors associated with large price movements in risk assessment~\cite{Poon2004}), and many others.
We provide a general null hypothesis test to determine whether coincidences are random in nature.

Throughout this work, we focus on the identification of noise transients (either terrestrial or instrumental) in ground-based GW interferometers such as the advanced LIGO~\cite{LIGO} and Virgo~\cite{Virgo} detectors.
GW interferometry provides a rich assortment of example analyses.
We refer to any short-duration ($\lesssim \mathcal{O}(1\,\mathrm{sec})$) transient within the instruments' sensitive frequency band ($20\,\mathrm{Hz}$ to a few kHz) as an \emph{event}.
When individual events can be accurately modeled and the transfer function between the event's source and the observed data is well understood, more complicated consistency checks can provide powerful tests of causal connections (e.g., \cite{Isi2018, Cornish2015, Kanner2016}).
This has been used to fantastic effect by combining accurate waveform predictions for GW signals expected from the coalescence of compact binary systems (e.g.,~\cite{Buonanno2009}) with precise knowledge of the detectors' interferometric response~\cite{Cahillane2017, Essick2019} in order to detect and infer the properties of some of the most extreme events in the universe~\cite{GWTC-1, GWTC-2}.

However, this is often not the case.
Whether direct measurements of the couplings between different sensors are too difficult to obtain, models of the transients are difficult to construct, or whether there are simply too many possible couplings between large numbers of sensors to accurately catalog, we are often faced with the general problem of determining the significance of coincidences between sets of discretized events without knowledge of their expected shapes.

Within GW astronomy, a variety of Event Trigger Generators (ETGs; typically distinguished by the wavelet transform employed, see \cite{Chatterji2004, godwin-thesis}) process discretely sampled time-series, or \emph{channels}, into tabular summaries of excess signal power that are well-localized in the time-frequency plane.
Each event is generally associated with a measure of the event's time, duration, and amplitude or signal-to-noise ratio ($\rho$, often representative of how rare the event is).
Within GW interferometers, there are typically $\mathcal{O}(10^4)$ separate channels that are sampled at frequencies fast enough to record noise within the detectors' sensitive band~\footnote{Current interferometers record in excess of $2\times10^5$ channels, but the majority are recorded a sample rates $\leq 16\,\mathrm{Hz}$}.
Astrophysical GW signals are recorded with the highest sensitivity by the designated ``GW channel'' (a measure of the antisymmetric motion of the L-shaped interferometer arms~\cite{Cahillane2017, Sun_2020}) but their passing may also be recorded by other auxiliary channels.
Conversely, terrestrial and instrumental noise recorded by auxiliary channels may be present in the GW channel.
Using auxiliary data to infer the non-astrophysical origin of noise artifacts in the GW channel can increase our confidence in the astrophysical nature of candidate events~\cite{Essick2020, Abbott2018}, but we generally do not have accurate models for many sources of noise, nor do we know the precise transfer functions between auxiliary channels and the GW channel (e.g., \cite{Cabero2019, Zevin2017}, although some monitors of the physical environment are given particular care~\cite{Effler:2014zpa}).
Again, without models that describe the relationship between events' waveforms (as they appear in multiple channels) and the GW channel, we are faced with the prospect of determining the significance of coincidences with only information about the events' time, $\rho$, and approximate rate at which they occur.

This problem has been studied within the context of GW interferometers' data quality in some detail.
Refs.~\cite{Essick2013} and~\cite{Smith2011} estimate the probability of observing a number of coincidences within a prespecified time window given an estimate of the rate of Poisson-distributed events.
That is, they construct a counting experiment to determine the number of GW noise artifacts that are within small time windows surrounding a series of events in an auxiliary channel, estimating the significance with the Poisson probability for the observed number of events given a point-estimate of the rate of events.
Ref.~\cite{Isogai2010} explores a similar approach, although their metric for the significance of coincidences is not derived from Poisson statistics.
These algorithms implement costly, direct searches over both the size of the time window and the rarity of the auxiliary artifacts about which the windows are placed.
They also intentionally learn to ignore many of the auxiliary channels through a supervised training regimen, extracting only a subset of auxiliary channels shown to correlate with noise in the GW channel.
If the selected auxiliary channel suddenly stops witnessing the source of noise (e.g., the sensor is unplugged), as occasionally happens, these algorithms lose all predictive power and must be retrained from scratch.
Additionally, they often do not clearly quantify null results, or the false dismissal probability.
That is, they do not provide a measure of how likely the observed data is if there is no observed coincidence within the time windows chosen \textit{a priori}.

More complicated approaches that utilize general machine learning algorithms have also been explored (e.g.,~\cite{Biswas2013, Essick2020}, see~\cite{Cuoco2020} for a review), but have been met with mixed success.
These may be improved with additional features based on our physical understanding of the situation.
For example, Ref.~\cite{Biswas2013} found that the most important features were the time separating auxiliary events from noise artifacts in the GW channel and how rare the auxiliary events were, implying that the algorithm spent most of its time learning how to determine the significance of coincidences.

This motivates the particular statistic investigated in this study.
In addition to providing additional features for existing algorithms, we find it generally useful to determine the false dismissal probability directly based on the proximity of the coincidence and the rarity of the events involved.
This can be determined from the receiver operating characteristic (ROC) curves for some existing algorithms, but the interpretation thereof may depend on the properties of the signals being sought.
That is to say, the false dismissal probability may depend on the selection criteria employed to identify the targeted events (loud noise artifacts may be fewer in number and easier to identify than more common quiet artifacts).
Essentially, we want to be able to quantify a statement like ``there is no important correlation or coincidence'' rather than stating that ``we did not find anything, but we do not know whether we would have found everything if anything was there.''

As such, we develop a specific null test in an attempt to directly quantify the significance of any putative coincidence.
In particular, we seek
\begin{itemize}
    \item{a quantifiable null test for the hypothesis that the time of interest is randomly drawn (i.e., uncorrelated with a stationary Poisson process),}
    \item{a more computationally efficient algorithm than a direct search over time-windows within counting experiments, and}
    \item{a more natural interpretation for our results when we have a single event of interest, rather than a large set of possible coincidences.}
\end{itemize}
We present a general null test under the assumption that the events are distributed according to stationary Poisson processes.
This improves upon previous approaches to estimate the significance of coincidences between gamma-ray bursts and GW events (Appendix B of \cite{Connaughton2016}), starting from first-principles distributions for noise artifacts and explicitly allowing for the possibility of multiple coincident artifacts for any particular time of interest.
This motivates our \emph{pointy statistic}, so named because it is a very pointed test of a specific null hypothesis with well-specified prior assumptions.
We then show its usefulness in a variety of situations, even when our motivating assumptions do not hold.

This paper is structured as follows.
Sec.~\ref{sec:formalism} motivates and derives the specific form of the pointy statistic.
Sec.~\ref{sec:applications} then presents several examples, each of which demonstrates a different way to utilize the pointy statistic.
We focus on examples from ground-based GW interferometry to demonstrate general techniques, but, as mentioned above, our techniques are immediately applicable within broader contexts.
We conclude in Sec.~\ref{sec:discussion}.

\section{Formalism}
\label{sec:formalism}

To formulate a general null test, we begin with a set of assumptions about individual channels.
Specifically, we assume that events in each channel are independent of events in other channels and that each set of events is distributed according to a stationary Poisson process (i.e., with a constant rate).
Although both of these assumptions are violated in many practical applications, they are generally good local approximations.
Additionally, they provide a starting point for more general analyses.

We are interested in the probability that a coincidence between an event and a (separate event at a) random time would be as close or closer than the observed separation.
That is to say, we want to know the probability that no events would occur within a window that is as wide as the observed coincidence.
To do this, we need to know how the events are distributed in time.
Specifically, we assume that the time between consecutive events $\Delta t$ in a single channel with a Poisson rate of $\lambda$ is distributed according to
\begin{equation}\label{eq:Delta t}
    p(\Delta t | \lambda) = \lambda e^{-\lambda \Delta t}.
\end{equation}
Given this assumption, we naturally define the probability of observing an event as close or closer than $\tau$ to an uncorrelated time-of-interest to be
\begin{equation}\label{eq:basic pvalue}
    P(\Delta t \leq \tau | \lambda) = 1 - e^{-2 \lambda \tau}
\end{equation}
where the factor of two is introduced by the fact that we search acausally (both backwards and forwards in time) to find the nearest event.

However, we do not observe the rate $\lambda$ directly, and instead estimate it from a counting experiment over a wider window.
We marginalize over the uncertainty in $\lambda$ such that
\begin{equation}\label{eq:marginalize}
    P(\Delta t \leq \tau | N, T) = \int d\lambda\, p(\lambda | N, T) P(\Delta t \leq \tau | \lambda)
\end{equation}
where
\begin{equation}\label{eq:lambda given N T}
    p(\lambda | N, T) = \frac{T}{N!} \left(\lambda T\right)^N e^{-\lambda T}
\end{equation}
which assumes a uniform prior on $\lambda$ and that the observed number of events ($N$) over the wider window ($T$) is Poisson distributed.
We then obtain
\begin{equation}
    P(\Delta t \leq \tau | N, T) = 1 - \left(1 + \frac{2 \tau}{T}\right)^{-(N+1)} .
\end{equation}
Alternatively, one could also assume a Jeffreys prior $p(\lambda) \propto \lambda^{-1/2}$, putting more weight on smaller rates \textit{a priori}.
However, we typically use large enough windows that $N \gg 1$, rendering the precise prior assumptions unimportant.
For that matter, simply using the point estimate $\hat{\lambda}=N/T$ within $P(\Delta t \leq \tau | \lambda)$ instead of marginalizing gives similar results, although there are differences in the tails.
Specifically, in the interesting limit of small $\tau$ ($\ll T$), we have
\begin{equation}\label{eq:unnorm pvalue}
    \lim\limits_{\tau \ll T} P(\Delta t \leq \tau | N, T) \sim 2\tau \frac{N+1}{T}
\end{equation}
whereas the
\begin{equation}
    \lim\limits_{\tau \ll T} P(\Delta t \leq \tau | \lambda = N/T) \sim 2\tau \frac{N}{T}.
\end{equation}
If $N \gg 1$, then $N+1 \approx N$ and the rate is measured well enough that marginalization has little effect.
However, the marginalized statistic (Eqn.~\ref{eq:unnorm pvalue}) will not vanish, even if $N=0$, as it accounts for the non-vanishing likelihood that the true rate is non-zero.

As a final practical consideration, we often limit the maximum $\tau$ considered to be within some relatively large coincidence window ($\tau \leq \mathcal{T} \ll T$) so that
\begin{equation}\label{eq:pvalue}
    P(\Delta t \leq \tau | \tau \leq \mathcal{T}, N, T) = \frac{1 - (1 + 2\tau/T)^{-(N+1)}}{1 - (1+2\mathcal{T}/T)^{-(N+1)}}
\end{equation}

Eqn.~\ref{eq:pvalue}, then, defines the \emph{pointy statistic} for a particular interval $\tau$ defined between a time-of-interest and single event.
It measures the probability of obtaining a coincidence as close or closer than the one observed assuming the events in the channel are distributed according to a stationary Poisson process and are uncorrelated with the time-of-interest.
If this probability (\pvalue) is small, then such coincidences are rare.
In this case, we may reject the null hypothesis that the event is uncorrelated with the time-of-interest, thereby inferring a causal connection.

Generally, the rate at which events occur within a channel also depends on their significance, or how loud they are, often characterized by a signal-to-noise ratio $\rho$.
More extreme excursions (larger $\rho$) are less likely than smaller ones.
That is to say, the Poisson rate depends on $\rho$, with $d\lambda/d\rho \leq 0$.
This presents a dilemma.
If the rate of quiet events is high enough, including the sea of quiet events within Eqn.~\ref{eq:pvalue} could drown out the significance of a loud coincident event.
We resolve this by considering multiple subsets of events corresponding to multiple thresholds on $\rho$, computing the pointy statistic for each subset.
To wit, we select subsets such that $\rho \geq \rho_\mathrm{thr}$ and then minimize the \pvalue~over $\rho_\mathrm{thr}$ such that
\begin{equation}\label{eq:pmin}
    P_\mathrm{min}(\Delta t \leq \tau) = \min\limits_{\rho_\mathrm{thr}} \left\{ P(\Delta t \leq \tau| \tau \leq \mathcal{T}, N(\rho \geq \rho_\mathrm{thr}), T) \right\}
\end{equation}
where $N(\rho \geq \rho_\mathrm{thr})$ is the number of events that satisfy $\rho \geq \rho_\mathrm{thr}$.
Typically, a handful of thresholds are chosen to reduce the computational cost of the direct minimization over $\rho_\mathrm{thr}$~\footnote{Note that other algorithms \cite{Essick2013, Smith2011, Isogai2010} must marginalize directly over both $\rho_\mathrm{thr}$ and a time window.}.
In this way, $P_\mathrm{min}$ is sensitive to the fact that rare, loud events (small $\lambda$) may constitute a more significant coincidence than quieter events (large $\lambda$) even if the louder events are further away (larger $\tau$).
On the other hand, quiet events (large $\lambda$) that are exceptionally coincident (small $\tau$) may be more significant than exceptionally loud events (small $\lambda$) that are very far away (large $\tau$).

It is also worth noting that estimates of the rate at different thresholds extracted from the same underlying set of events will be correlated~\cite{Lynch2018}.
As such, a more complete analysis should include this correlated uncertainty during the minimization over $\rho_\mathrm{thr}$ or otherwise account for the correlated knowledge of coincidences' significance at different thresholds.
However, direct minimization works well in practice, and we leave further theoretical exploration to future work.

Nonetheless, minimization implies that our \pvalues~may no longer be distributed as one would na{\"i}vely expect when selecting random times.
That is to say, they are not necessarily distributed uniformly between $0$ and $1$.
While this slightly complicates the statistical interpretation of $P_\mathrm{min}$, it does not pose a significant problem; we directly measure the distribution of $P_\mathrm{min}$ from randomly selected times to establish rigorous false alarm probabilities (FAPs).

Equipped with Eqn.~\ref{eq:pmin}, we evaluate $P_\mathrm{min}$ at every point in a regularly sampled time-series (see below), obtaining results like those in Fig.~\ref{fig:timeseries}.
We additionally impose a minimum $\tau$ based on a fraction ($F$) of the most significant event's durations ($\Delta$).
This approximates our uncertainty in the individual events' central times, as these are not known precisely but should have an uncertainty that scales with their durations.
Specifically, we assign a value
\begin{equation}\label{eq:tau}
    \tau = \min_{i \in N(\geq \rho_\mathrm{thr})} \{\mathrm{max}\{|t-t_i|,\ F\Delta_i\}\}
\end{equation}
to each time $t$ in the time-series, associating it with the nearest event above $\rho_\mathrm{thr}$, and then compute $P_\mathrm{min}$ via Eqn.~\ref{eq:pmin} while marginalizing over the unobserved Poisson rate.
This produces small $P_\mathrm{min}$ near the locations of events that decay approximately exponentially.
Minimization over different $\rho_\mathrm{thr}$, and therefore different estimates of the Poisson rate, produces peaks of different widths in the time-series.
Common, quiet events have relatively narrow peaks as they occur at a high rate and therefore their significance drops off quickly.
Loud, rare events form much broader (possibly deeper) peaks because they remain significant even with broader coincidence windows.

\begin{figure*}
    \includegraphics[width=\textwidth]{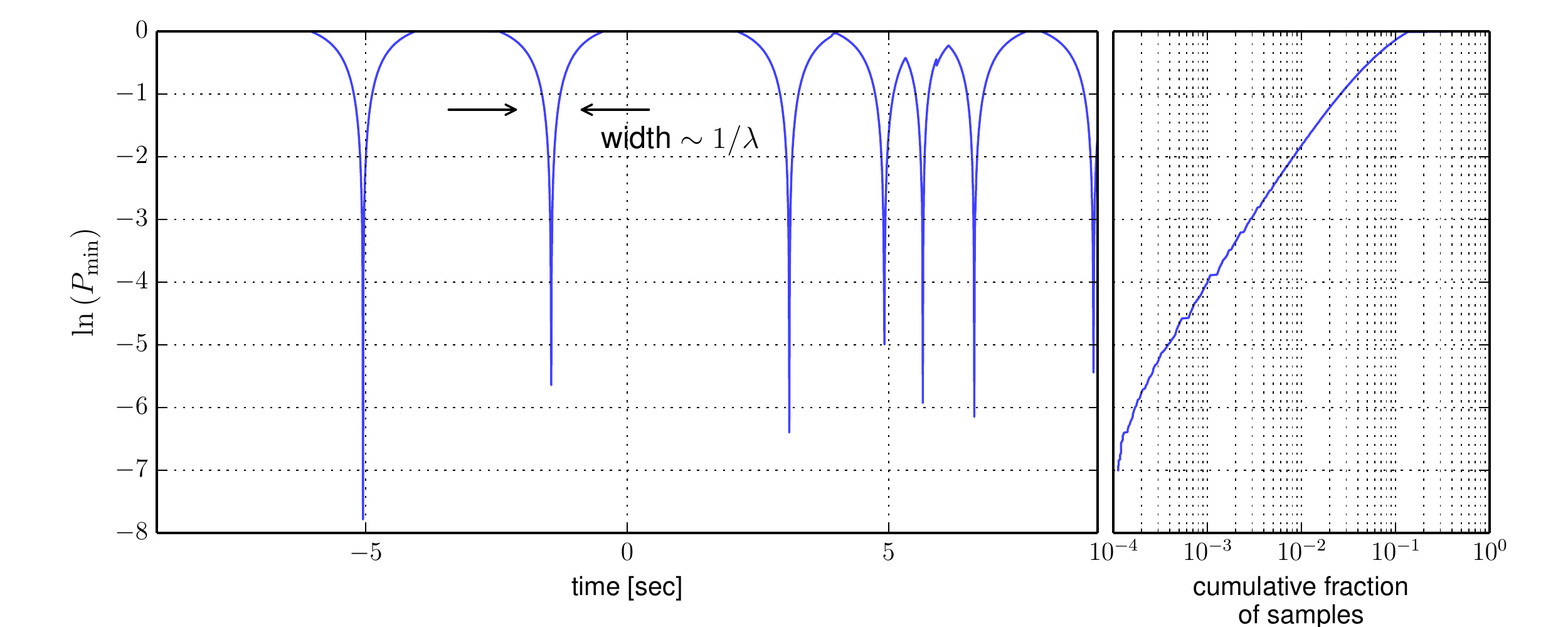}
    \caption{
        (\emph{left}) An example $P_\mathrm{min}$ time-series for a seismic isolation control channel in the LLO x-arm's input test mass's suspension system (\result{L1:ISI-ITMX\_CDMON\_ST2\_V1\_V\_IN1\_DQ}) between 8 and $128\,\mathrm{Hz}$, centered at \result{22:35:30 GMT on Tue Sep 3, 2019}.
        The sequence of peaks in the time-series correspond to a sequence of events in this channel, and the width of each peak roughly corresponds to the estimated Poisson rate.
        (\emph{right}) A cumulative histogram of $P_\mathrm{min}$, from which we can directly measure the probability of observing smaller $P_\mathrm{min}$ by random chance (i.e., the false alarm probability).
    }
    \label{fig:timeseries}
\end{figure*}

Our analyses focus on different ways to combine $P_\mathrm{min}$ from a variety of auxiliary channels and times.
Typically, we assume statistical independence between the different auxiliary channels within na{\"i}ve Bayes classification schemes, finding that the assumption of independence renders analyses tractable while still performing well.
Specifically, we generate $P_\mathrm{min}$ separately for a collection ($\mathcal{S}$) of separate channels and/or times, combining them under the assumption of statistical independence so that
\begin{equation}\label{eq:pjoint}
    P_\mathrm{joint} = \prod\limits_{i \in \mathcal{S}} P_\mathrm{min}^{(i)} .
\end{equation}
We refer to this practice of combining $P_\mathrm{min}$ into $P_\mathrm{joint}$ as \emph{stacking}.
Sec.~\ref{sec:applications} describes a few ways in which we select sets $\mathcal{S}$ that are of physical relevance within GW data analysis.

\section{Applications}
\label{sec:applications}

We demonstrate the pointy statistic with two examples.
We will first establish channel \emph{safety} in interferometric GW detectors in Sec.~\ref{sec:single channel, multiple times}, meaning whether veto conditions based on auxiliary channels could systematically reject real GW signals.
This is a key aspect of any veto condition in GW transient searches.
This application explores situations where we can identify repeated experiments, implying that we can stack $P_\mathrm{min}$ for each channel from different times to learn more about channels separately.
Sec.~\ref{sec:multiple channels, single time} presents a second example of our pointy statistic where it constructs a veto condition based on the simultaneous use of information from multiple safe auxiliary channels.
This allows us to investigate particular times in GW detectors and quantify the possible correlations between noise in the GW channel and auxiliary sensors.

\subsection{Veto Safety in Gravitational-Wave Searches}
\label{sec:single channel, multiple times}

We first focus on establishing channel safety in interferometric GW detectors, and, more specifically, the application of veto conditions in GW transient searches.
GW interferometers, like advanced LIGO and Virgo, record not only the astrophysical GW signal but also a plethora of auxiliary channels that monitor the instruments' states and their physical environments~\cite{Mueller:2016hex, Matichard:2015eva, Staley:2015nie, Rollins:2016hlk, Effler:2014zpa}.
Information from these auxiliary channels can be used as vetoes for putative astrophysical events.
That is to say, noise artifacts that pollute the GW channel may also appear in auxiliary channels.
However, although the main GW channel records the astrophysical signal with the highest sensitivity, some auxiliary channels may also witness GW signals either inadvertently or because they are part of the detector's feedback control scheme.
This can confuse the veto inference; if veto conditions are based on channels that have some sensitivity to GW signals, then these conditions may systematically reject real astrophysical events~\footnote{It is worth noting that the concept of safety may be defined in terms of the raw auxiliary channel time-series themselves or in terms of the events produced with a specific ETG. The former is often what is sought, but we are forced to use a single ETG due to computational limitations and must settle for some notion of the latter.}, the most extreme example of which would be constructing a veto based on the GW channel itself (e.g.,~\cite{Zackay:2019}).
In many analyses then, it is of the utmost importance to determine which auxiliary channels witness GW signals and are therefore \emph{unsafe} for veto conditions.
Most approaches to data quality and vetoes limit themselves \textit{a priori} to the subset of auxiliary channels that can be demonstrated to be safe ~\cite{Essick2020}.

Due to the complexity of kilometer-scale interferometers, it is not possible to directly measure every coupling between the over $2\times10^5$ auxiliary channels at each interferometer (or even the $\mathcal{O}(10^4)$ sampled faster than $16\,\mathrm{Hz}$).
Therefore, we determine correlations probabilistically.
This has historically been done with a set of transient injections performed in hardware by directly manipulating the instruments.
Typically, an external force of known strength is applied to the interferometer's test masses, inducing differential arm motion that is read out analogously to an astrophysical signal.
We additionally demonstrate how actual astrophysical events provide another method to independently ``inject'' signals into the interferometers, allowing one to establish auxiliary channel safety.

\subsubsection{Safety Studies with Hardware Injections}
\label{sec:hardware_injections}

Hardware injections in GW interferometers directly manipulate the length of one of the detector's arms, producing a change in the differential arm length that mimics the effects of a GW signal.
If the injected signal appears in an auxiliary channel, then it is also possible for a real GW to appear in that channel, and the channel is declared unsafe.
Furthermore, couplings may depend on both the GW signal's frequency (e.g., different seismic isolation subsystems intentionally have very different responses at low and high frequencies) and its amplitude (a quiet GW signal may not show up significantly louder than the Gaussian noise in an auxiliary channel, and therefore may not produce a trigger with a particular ETG).
As we can control the number and the properties of the hardware injections, but do not know the transfer functions between the injection site and all auxiliary channels, we adopt a statistical approach to safety.
In this situation, we are interested in the behavior of each auxiliary channel separately, which provides the perfect example of how to combine independent, identically distributed trials with the pointy statistic.
As such, we produce a time-series representing the minimized pointy statistic (Eqn.~\ref{eq:pmin}), computing a joint \pvalue~(Eqn.~\ref{eq:pjoint}) by stacking the individual $P_\mathrm{min}$ associated with the time of each injection in a sequence ($\mathcal{S}$).

We summarize results obtained with hardware injections performed in the LIGO Livingston Observatory (LLO) during LIGO's third observing run (O3)~\cite{LLOalog}.
We injected sine-Gaussian signals as a general proxy for GW transient events that are compact in the time-frequency plane (spanning a short duration and limited to a narrow frequency bandwidth), allowing us to isolate possible couplings at different frequencies.
The injections varied in amplitude, corresponding to expected $\rho$ from 15 to 500 and with central frequencies ($f$) between $20\,\mathrm{Hz}$ and $700\,\mathrm{Hz}$.
Each $f$-$\rho$ combination was repeated 3 times, with each injection separated by $5\,\mathrm{sec}$.
Over $5,500$ auxiliary channels at LLO were analyzed for all injections using an ETG based on a dyadic Haar wavelet decomposition (KleineWelle: KW~\cite{Chatterji2004, kleine-welle}).

Following the formalism of Sec.~\ref{sec:formalism}, we used a time window ($T\sim5000\,\mathrm{sec}$) much larger than the injection durations (each injection lasted for $\sim 50\,\mathrm{msec}$ and the sequence of 84 injections spanned $435\,\mathrm{sec}$) and the number of observed events above a set of thresholds ($N(\rho\geq\rho_\mathrm{thr})$) in each auxiliary channel separately to marginalize over the Poisson rate ($\lambda$) for each threshold (Eqn.~\ref{eq:marginalize}) before minimizing the $P_\mathrm{min}$ over $\rho_\mathrm{thr}$ (Eqn.~\ref{eq:pmin}) to identify the most significant coincident event.
This procedure was conducted repeatedly to build up a $P_\mathrm{min}$ time-series sampled at $128\,\mathrm{Hz}$ spanning the full $5000\,\mathrm{sec}$ window.
The $P_\mathrm{min}$ corresponding to the times of each triplet of hardware injections with the same $f$ and $\rho$ were then extracted via linear interpolation, constituting a set ($\mathcal{S}$) that we stack to compute $P_\mathrm{joint}$ (Eqn.~\ref{eq:pjoint}).

\begin{figure}
    \begin{center}
        \includegraphics[width=1.0\columnwidth, clip=True, trim=0cm 1.50cm 0.00cm 0.15cm]{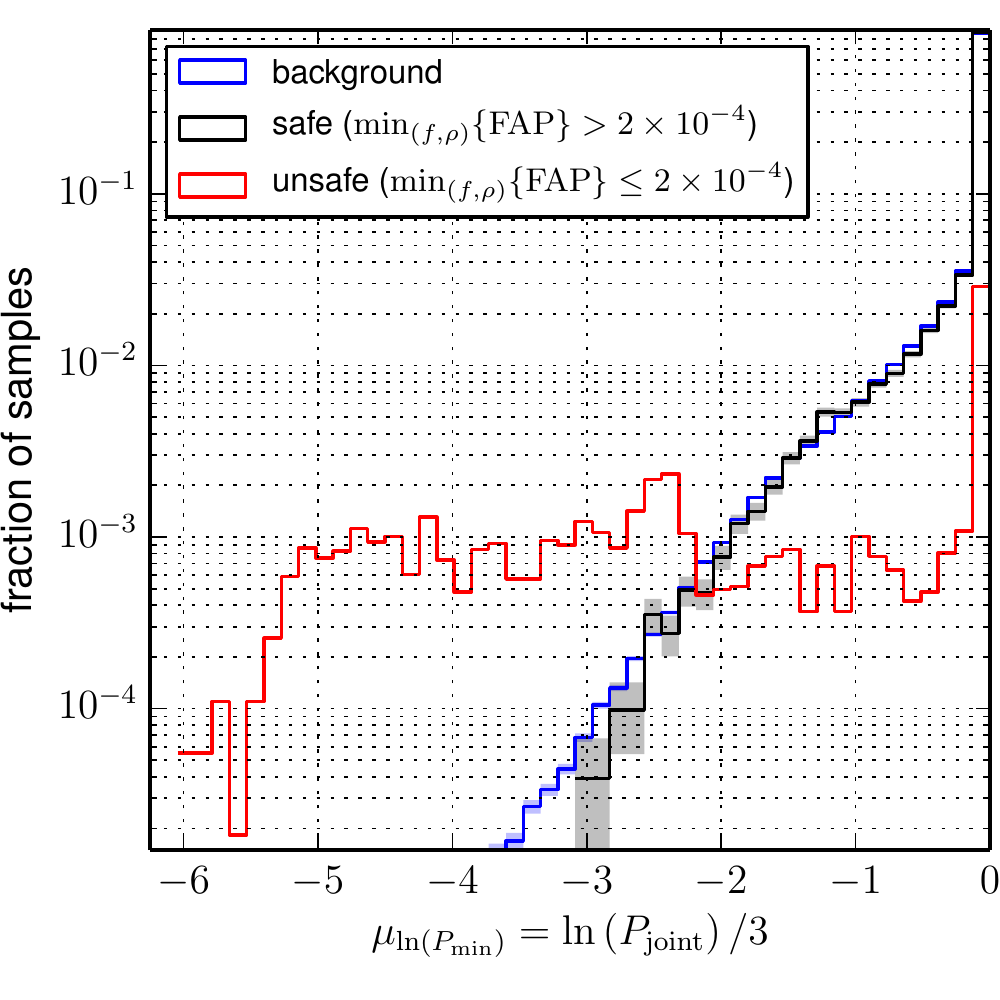}
        \includegraphics[width=1.0\columnwidth, clip=True, trim=0cm 0.40cm 0.00cm 0.15cm]{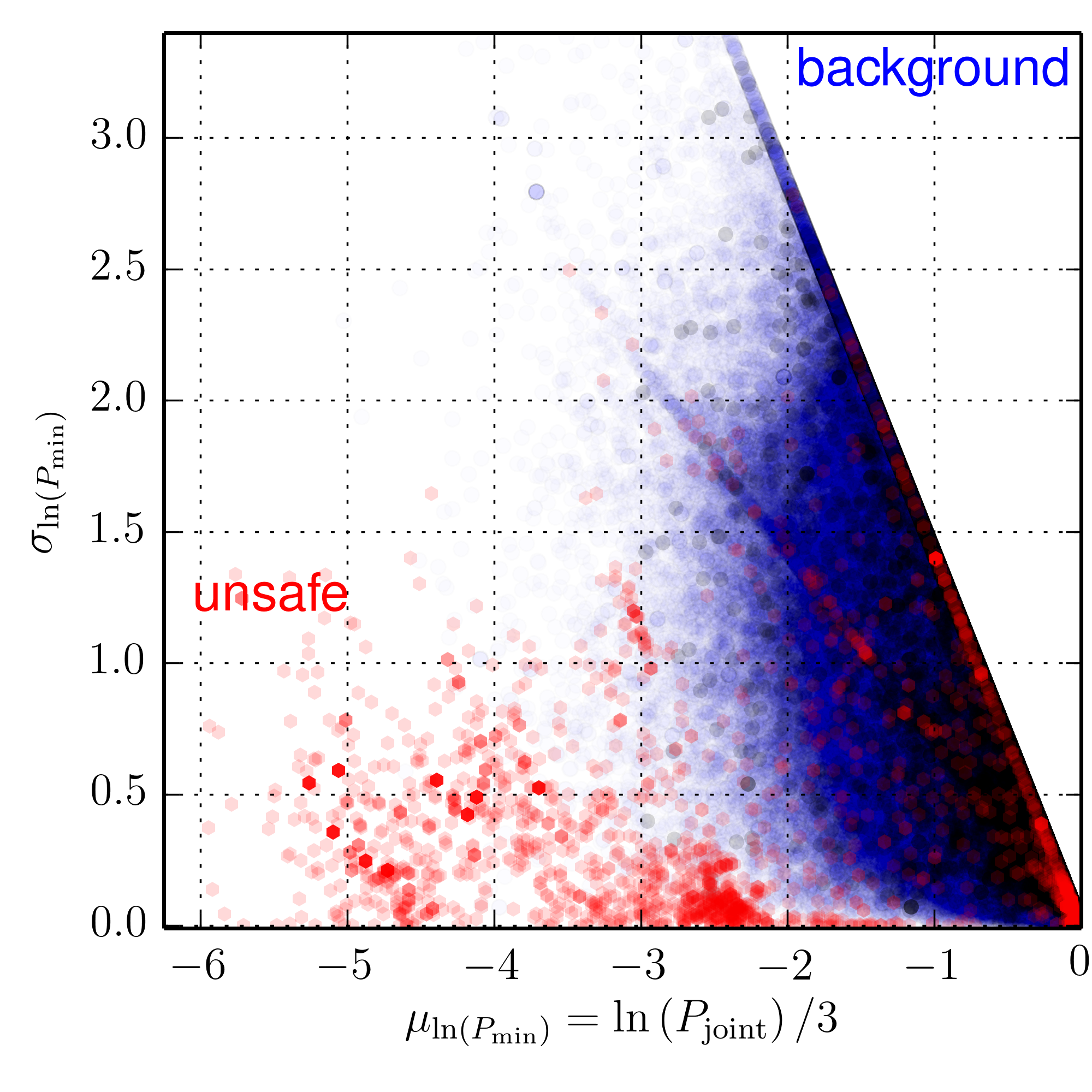} 
    \end{center}
    \caption{
        Summary of results from hardware injections at LLO.
        (\emph{top}) Distributions of $P_\mathrm{joint}$ for channels declared unsafe (\emph{red}), channels declared safe (\emph{black}), and an estimate of the expected background (\emph{blue}).
        Shaded bands represent 1-$\sigma$ uncertainty estimates from counting statistics, showing good agreement between the background and safe distributions.
        We present distributions over all ($f$, $\rho$)-injection triplets, and safety was determined by the minimum FAP observed for each channel over all triplets.
        The unsafe distribution extends to $P_\mathrm{joint} \sim 1$ because some channels that are unsafe at low frequencies and large $\rho$ are safe at high frequencies and/or small $\rho$.
        (\emph{bottom}) Joint distributions of $P_\mathrm{joint}$ and the standard deviation of $P_\mathrm{min}$ between the three repeated injections in each triplet ($\sigma_{\ln(P_\mathrm{min})}$).
        Although there is clearly additional structure in the two-dimensional distributions for both the unsafe (\emph{red hexagons}) and background (\emph{blue circles}), we find that safety criteria based solely on $P_\mathrm{joint}$ accurately captures all the relevant information.
    }
    \label{fig:hardware injections}
\end{figure}

Fig.~\ref{fig:hardware injections} demonstrates the observed distributions of $P_\mathrm{joint}$ for triplets of hardware injections and randomly selected sets of 3 times not associated with hardware injections that empirically determine the pointy statistic's background distribution.
Specifically, we select 3 random times repeatedly for each channel, computing $P_\mathrm{joint}$ for each random triplet.
The total ensemble then empirically determines the expected FAP at each $P_\mathrm{joint}$.
The pointy statistic also normalizes all auxiliary channels so that we can fairly compare the significance of coincidences in each.
\textit{De facto}, this is done by not only considering $\Delta t$ but also by incorporating knowledge of the Poisson rate of accidental coincidences.
Indeed, for this reason, estimates of the background distributions for individual channels closely resemble our background estimate that averages over all channels.
We note that we expect $\mathrm{FAP} \sim P_\mathrm{min}$ by construction, as Fig.~\ref{fig:timeseries} demonstrates.

If the interferometric data was truly Poisson distributed and the repeated injections were exactly identical and independent, then $P_\mathrm{joint}$ should be a sufficient statistic.
That is to say, no additional information would be available to an analysis that considered the set of 3 $P_\mathrm{min}$ for each injection triplet compared to one that only considered the corresponding $P_\mathrm{joint}$.
Fig.~\ref{fig:hardware injections} shows projected histograms of $P_\mathrm{joint}$ for both the hardware injection triplets and for the background distribution.
Indeed, we see that most of the information available is captured in this statistic.
However, because the motivating assumptions behind the pointy statistic may be violated in practice (see Appendix~\ref{sec:assumptions}), we additionally consider a higher-dimensional inference.
Specifically, we consider the joint distribution between $P_\mathrm{joint}$ and $\sigma_{\ln(P_\mathrm{min})}$ (the standard deviation obtained from the triplet of the natural logarithm of $P_\mathrm{min}$).
For deterministic couplings, we would expect nearly identical $P_\mathrm{min}$ for each injection and therefore small $\sigma_{\ln(P_\mathrm{min})}$.
However, for rare but nonetheless accidental coincidences, we expect large $\sigma_{\ln(P_\mathrm{min})}$ as rare coincidences that are not likely to be repeated.
Indeed, this is exactly the behavior we see in the background distribution ($\ln(P_\mathrm{joint}) \propto \sigma_{\ln(P_\mathrm{min})}$).

It is noteworthy that, even though real noise artifacts are not perfectly Poisson distributed, the information available in the joint two-dimensional inference is nonetheless captured by the one-dimensional inference over only $P_\mathrm{joint}$ with the additional caveat that a channel that is found to be unsafe for any $f$-$\rho$ pair is declared unsafe for all $f$-$\rho$ pairs~\footnote{While our analysis identifies both the frequency and the amplitude at which auxiliary channels become unsafe, current applications of safety information do not account for these dependencies. We therefore adopt the logic that channels are safe if and only if they are shown to always be safe.}.
That is to say, unsafe channels identified in the two-dimensional analysis for a single $f$-$\rho$ pair but were safe in the one-dimensional analysis of the same $f$-$\rho$ pair were either associated with different frequency bands in the ETG (KW divides ``raw channels'' into several overlapping bandpasses, producing a set of events for each) that witnessed an injection less efficiently or with frequency-dependent transfer functions between the injection site and the auxiliary channel, which typically caused injections to appear with lower significance at higher frequencies.
Both effects meant that the significance of coincidences was compared against an elevated rate of lower-$\rho$ background artifacts in each channel, thereby reducing the $P_\mathrm{joint}$ below the one-dimensional analysis's threshold.
However, because the injection set spanned a large range of frequencies, these channels were always identified as unsafe by either a separate injection triplet or a separate KW bandpass produced from the same raw channel.

Another important feature of our analysis is the direct assignment of a FAP to each channel.
While final lists of safe and unsafe channels were still collated based on a hard threshold, our analysis allows that threshold to be chosen based on the expected number of false positives rather than an arbitrary detection statistic without an immediate physical interpretation.
Within this study, we divided the channels into three sets: those found to be confidently unsafe ($\mathrm{FAP} \lesssim 2\times 10^{-4}$), those found to be confidently safe ($\mathrm{FAP} \gtrsim 2\times10^{-3}$), and suspicious channels in between ($2\times10^{-4} \lesssim \mathrm{FAP} \lesssim 2\times10^{-3}$).
These thresholds were chosen so that the expected number of false positives was \result{$\sim 1$} for the confidently unsafe class and $\lesssim 10$ for the suspicious class based on the fact that we analyzed $\approx 5,500$ channels.
Indeed, inspecting the identified channels' behavior by hand, we found \result{$1.9\pm1.1$} false positives (mean and standard deviation between different injection triplets) within the channels identified with $\mathrm{FAP} \lesssim 2\times10^{-4}$ and \result{$10.3\pm3.4$} channels with $2\times10^{-4} \lesssim \mathrm{FAP} \lesssim 2\times10^{-3}$, in agreement with expectations for a basic counting experiment.
Fig.~\ref{fig:hardware injections} shows the suspicious and confidently safe sets together, and we see that it closely follows the background distribution.

This analysis identified all \result{69} channels considered unsafe throughout O3 based on statistical evidence from hardware injections, including all channels known to be unsafe \textit{a priori} as well as several that were not.
Additional channels were also declared unsafe out of an abundance of caution because they either witnessed similar physical signals to channels that were identified as statistically unsafe (\result{64} channels), in which case small changes to physical couplings could render the uncorrelated channels unsafe, or had historically been declared unsafe even if they no longer could be shown to correlate with hardware injections (\result{45} channels).
Furthermore, the consistency of the observed rate of false positives with expectations based on our FAP thresholds suggests that our analysis was able to cleanly separate the entire unsafe channel population, implying that all statistically unsafe channels were robustly identified.
If this was not the case, then we would expect an excess of suspicious channels.
Indeed, this emphasizes the pointy statistic's utility as it allows a way to efficiently extract all relevant information about coincidences, even when its motivating assumptions are not perfectly satisfied, and our ability to extract more information about individual channels by examining repeated experiments simultaneously.

\subsubsection{Safety Studies with Astrophysical Events}
\label{sec:astrophysical injections}

While hardware injections allow us to perform repeated experiments, they inevitably require the interferometers to be manipulated in ways they would not be by a \textit{bona fide} GW signal.
For example, the channels that record the excitations introduced into the interferometer will obviously correlate with the injected signals, and therefore hardware injections cannot be used to determine the safety of the excitation channels.
If one can inject signals into the interferometer in multiple ways (e.g., by driving the length of each arm separately), then one could perform multiple sets of injections, using each to cross-check the channels used during the other set.\footnote{We expect excitation channels to be safe \textit{a priori} in that the presence of coincident transients in the excitation channels would call into doubt the astrophysical nature of any GW candidate. Nonetheless, one may want to confirm this out of an abundance of caution.}
However, we present an alternative that uses the ``perfect'' injections available through confident GW detections.

As of the time-of-writing, the advanced LIGO and Virgo detectors have published several dozen confident detections of binary black hole (BBH) coalescences~\cite{GWTC-1, GWTC-2}.
The astrophysical nature of these events is not in doubt, and therefore we can exploit them as natural experiments to determine which auxiliary channels witness real GW signals.
Previous work has shown how known astrophysical sources can be used to calibrate networks of GW interferometers~\cite{Essick2019}, and herein we extend this approach to auxiliary channel safety.

However, using astrophysical events to determine veto safety presents some limitations.
High-mass BBH mergers detectable in the LIGO interferometers are characterized by rapid frequency evolution and only extend up to relatively low frequencies ($\mathcal{O}(100\,\mathrm{Hz})$).
Although these astrophysical signals do not require direct manipulation of the interferometers, they also cannot be repeated.
Each BBH merger comes from a separate astrophysical system with different component masses and spins, meaning that the maximum frequency reached during the coalescence and the signal amplitude will differ for each system.
As channel safety is known to depend on the signal frequency and amplitude, we cannot stack multiple astrophysical events without extreme care.

\begin{figure}[!htpb]
    \begin{center}
        \includegraphics[width=1.0\columnwidth, clip=True, trim=0cm 1.50cm 0.00cm 0.15cm]{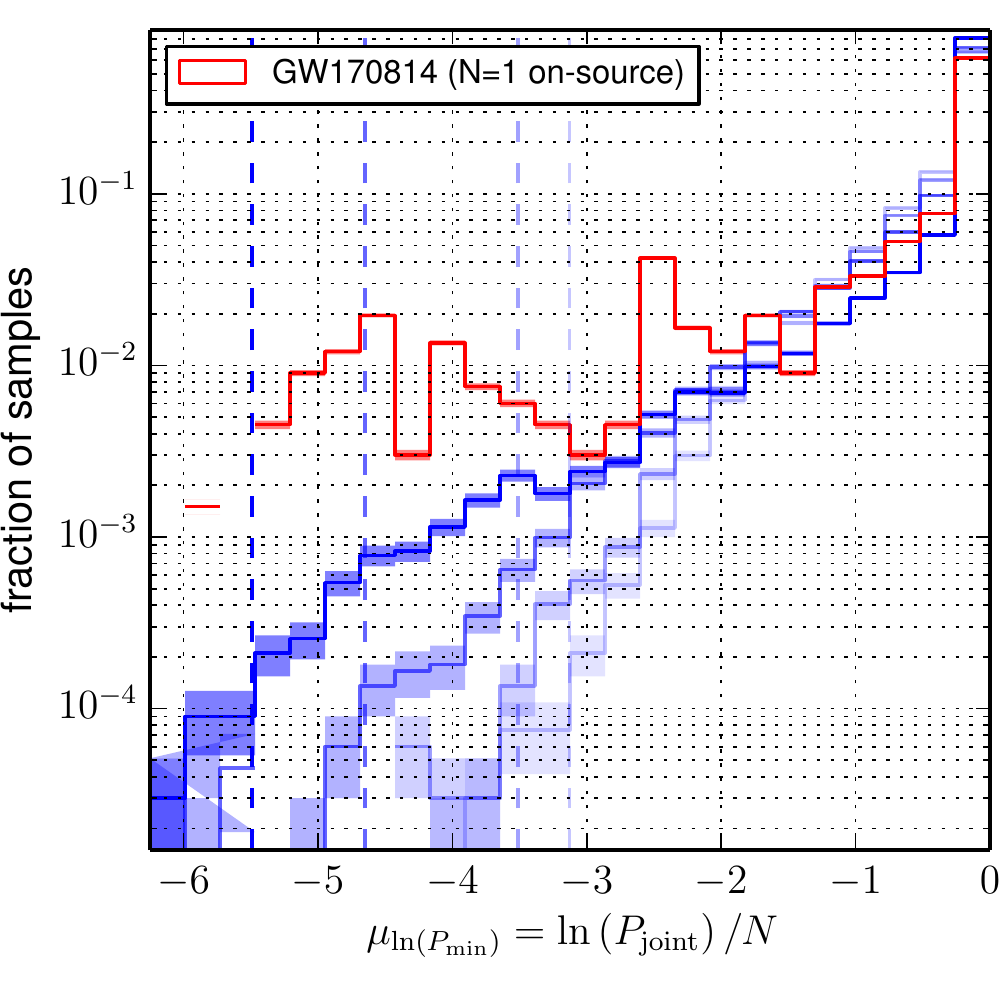}
        \includegraphics[width=1.0\columnwidth, clip=True, trim=0cm 0.40cm 0.00cm 0.15cm]{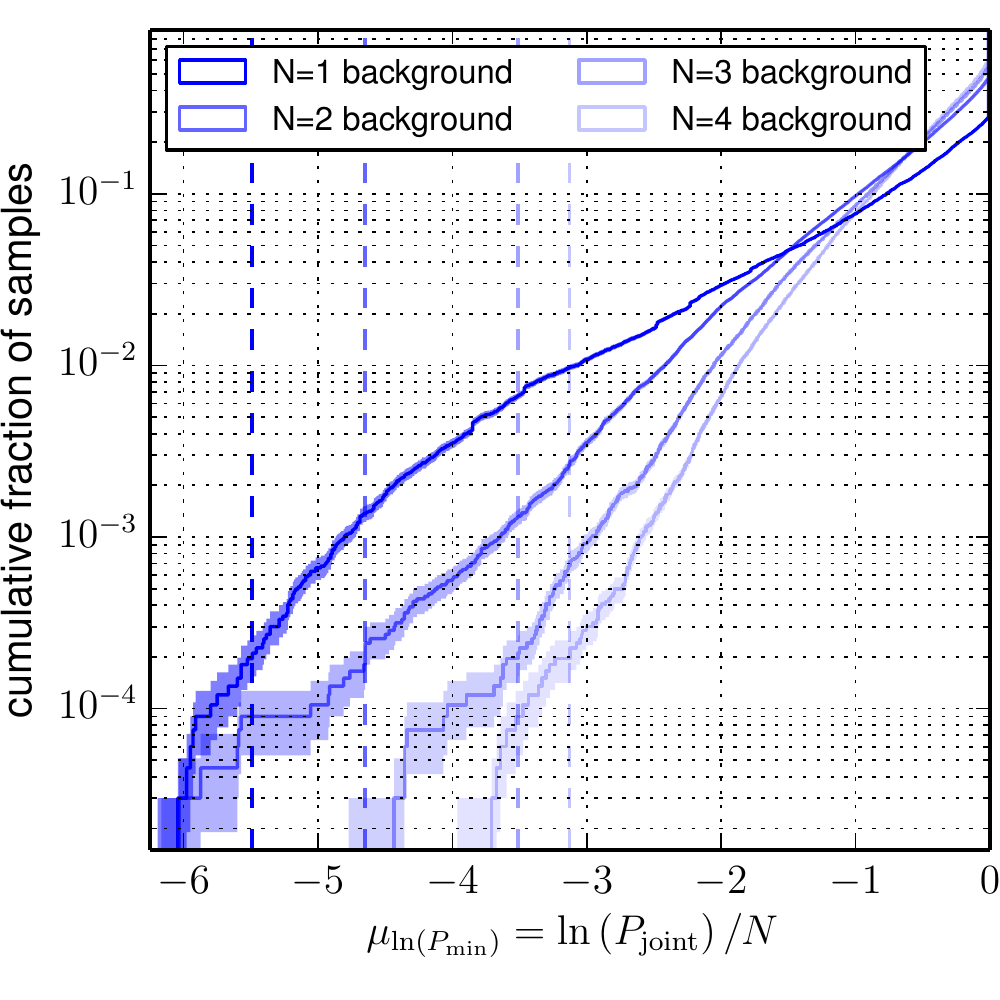}
    \end{center}
    \caption{
        (\emph{top}) $P_\mathrm{joint}$ distributions of background estimates obtained by stacking different numbers of random times ($N$), with the observed distribution from GW170814 superimposed.
        Vertical lines denote the approximate value of $P_\mathrm{joint}$ required to confidently detect an unsafe channel ($\mathrm{FAP} \lesssim 2\times10^{-4}$) given the large number of auxiliary channels considered.
        (\emph{bottom}) Cumulative $P_\mathrm{joint}$ distributions for the same background estimates.
        While GW170814 clearly shows an excess of channels with small \pvalues, we find that we need to combine at least three repeated observations before the corresponding $P_\mathrm{joint}$ would be clearly separated from the background.
    }
    \label{fig:astrophysical_safety}
\end{figure}

Nonetheless, we examine an individual confident BBH detection to establish channel safety as a proof-of-principle.
We analyze GW170814~\cite{GW170814, GW170814-aux-data} in LLO, where it was detected with a $\rho = 13.7$, much less than the $\rho\sim500$ achieved at similar frequencies with hardware injections.
Again, we processed approximately 5,500 auxiliary channels with KW, each sampled faster than $16\,\mathrm{Hz}$, generating pointy time-series from the resulting sequences of events.
We again estimated the background distribution of $P_\mathrm{min}$ from accidental coincidences by randomly selecting times from the $\sim5000\,\mathrm{sec}$ surrounding GW170814 for each channel.
We additionally require that these random times be at least $1\,\mathrm{sec}$ away from loud noise artifacts in the GW channel to avoid contaminating the background distribution with auxiliary channels that may witness such noise.
Fig.~\ref{fig:astrophysical_safety} shows the result.

We see a clear excess at small $P_\mathrm{min}$ within the on-source distribution.
However, the background distribution for a single random time has significant support down to even lower $P_\mathrm{min}$.
This implies that the FAP associated with the unsafe channels ($P_\mathrm{min} \lesssim e^{-4}$) is as high as $5\times10^{-3}$, implying we would be forced to accommodate $\sim25$ false positives if we were to identify all channels known to be unsafe in this way.
Therefore, we conclude that it is unlikely that a single BBH will produce on-source coincidences significant enough to overcome the large trials factor from the $\mathcal{O}(10^4)$ auxiliary channels considered.
The confident identification of channel safety with low contamination therefore requires repeated hardware injections.
These should be conducted regularly throughout observing runs, which has not been the case to-date during the advanced detector era.

Fig.~\ref{fig:astrophysical_safety} goes further, estimating the background that would be obtained by stacking more and more events.
We see that, if we require $\mathrm{FAP} \lesssim 10^{-4}$ at $P_\mathrm{joint} \sim e^{-4}$ to limit the expected number of false positives to $\lesssim 1$, we need to stack at least 3 repeated experiments.
The hardware injection campaign described in Sec.~\ref{sec:hardware_injections}, then, performed the minimum number of injections needed to cleanly separate safe and unsafe channels.

\subsection{Glitch Identification via Na\"{i}ve Bayes Classification}
\label{sec:multiple channels, single time}

Sec.~\ref{sec:single channel, multiple times} described how to determine which auxiliary channels are safe, and therefore can be used within veto conditions.
We now demonstrate the construction of such a veto condition tailored to a specific class of non-Gaussian noise artifact, colloquially referred to as \emph{whistles}.
While pointy analyses have also targeted other classes of noise, most notably a study of \emph{blip} glitches during LIGO's first observing run (O1), these have since been investigated in some detail in other work~\cite{Cabero2019}.
We focus on whistles to demonstrate unique advantages of the na\"{i}ve Bayes classification with the pointy statistic and to avoid repeating conclusions already presented elsewhere.

Our specific use-case was motivated by a small statistical excess of foreground events in a search for unmodeled GW transients during the second observing run (O2)~\cite{Lynch2017, O2burst}.
While no single event in that search was significant enough to claim a detection, a collection of events in the tail of the distribution constituted a small excess above what was expected, which prompted further follow-up.
Inspecting a collection of the most significant foreground events, we determined that the tail was mostly comprised of either blip or whistle glitches.
While it is known that the majority of blip glitches do not have clear auxiliary witnesses (see, e.g.,~\cite{Cabero2019}), whistles typically do have clear auxiliary witnesses and should be straightforward to veto.

As such, we analyzed the full set of auxiliary channels surrounding four whistles from the tail of the background distribution.
For each whistle, we identified the set of auxiliary channels found to be in coincidence with the background event with $P_\mathrm{min} \leq 3\times10^{-2}$ (the precise value has little impact), corresponding to a relatively high FAP ($\gtrsim 2\%$) and a large number of expected accidental coincidences given the large number of channels analyzed.
This procedure identified \result{$\sim 130$} channels that are of interest around each whistle (for some as few as \result{87}, for others as many as \result{208}), most of which were likely accidental coincidences.
While some individual auxiliary witnesses (typically angular sensing and control (ASC) and length sensing and control (LSC) channels) produced $P_\mathrm{min}$ small enough to possibly veto these background events separately~\footnote{Some individual channels produced \result{$P_\mathrm{min} \leq 10^{-3}$, implying $\mathrm{FAP} < 10^{-4}$} (see the $N=1$ background estimate Fig.~\ref{fig:astrophysical_safety}) and an expected number of accidental coincidences $\lesssim 1$}, we note that there was also an excess of channels with slightly larger $P_\mathrm{min}$ beyond what one would expect from a basic counting experiment (c.f., sanity checks in Sec.~\ref{sec:hardware_injections}).
As such, we were additionally interested in using information from all auxiliary witnesses simultaneously rather than only the individual channels that were most significant.
In this situation, the set $\mathcal{S}$ in Eqn.~\ref{eq:pjoint} represents a collection of separate auxiliary channels evaluated for a single time-of-interest, rather than $P_\mathrm{min}$ from a single channel evaluated at multiple times.

Using all channels identified as possibly significant around each whistle, we construct a na\"{i}ve Bayes statistic by assuming the channels were independent (though they likely are not) and multiplying their individual $P_\mathrm{min}$ to obtain $P_\mathrm{joint}$.
While many of the channels were accidental coincidences, if a subset of auxiliary channels did indeed witness the noise source, then their individual significances should stack together to provide a more confident veto than any individual channel would produce.
Indeed, this is exactly the behavior we observe.

\begin{figure}
    \begin{center}
        \includegraphics[width=1.0\columnwidth, clip=True, trim=0.15cm 0.10cm 0.27cm 0.40cm]{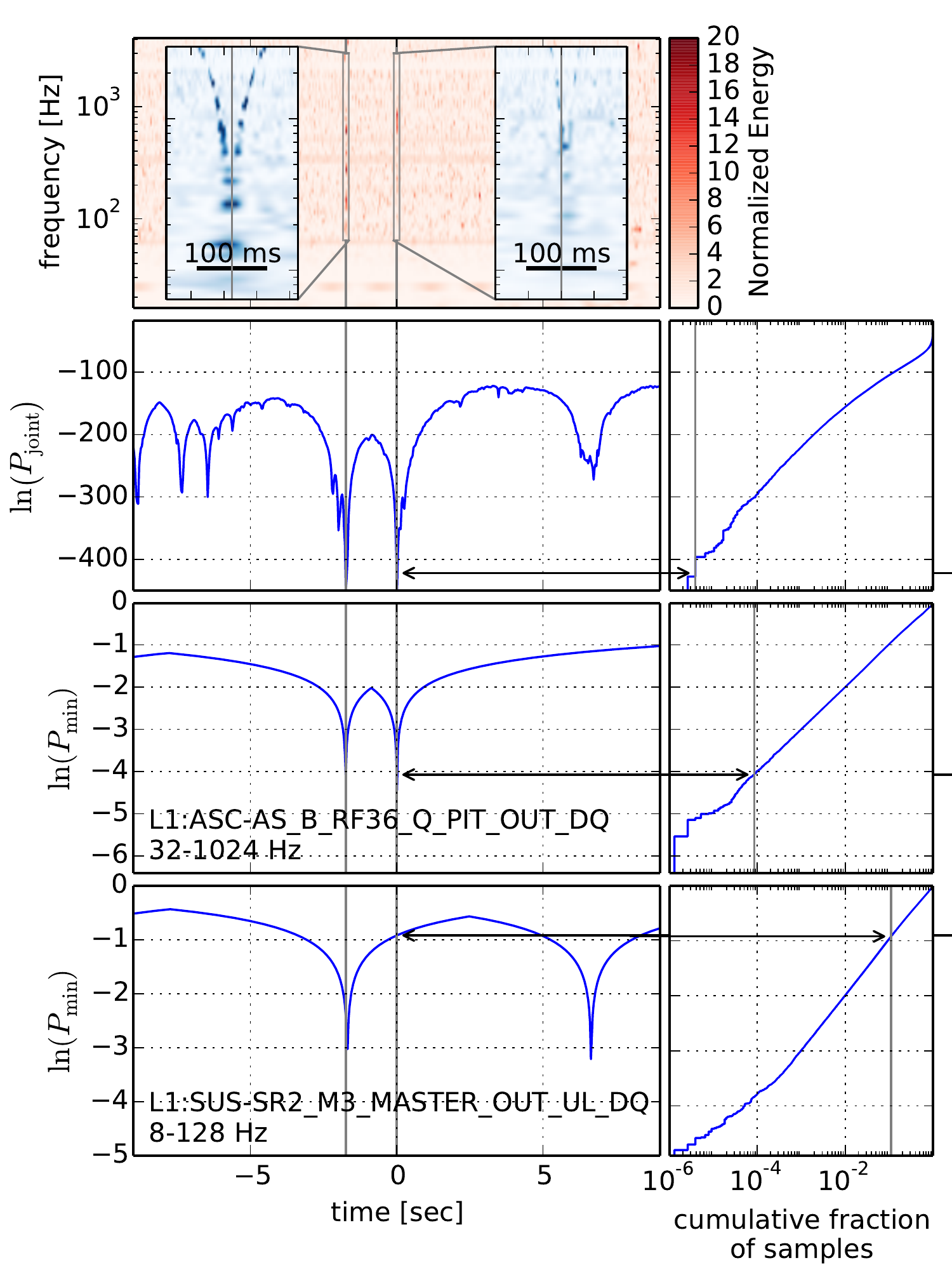}
    \end{center}
    \caption{
        An example whistle glitch identified in the background of a search for unmodeled transients during O2.
        (\emph{top}) A time-frequency representation~\cite{Chatterji2004} of \result{$20\,\mathrm{sec}$} of the strain data from LLO, with insets highlighting the noise identified by $P_\mathrm{joint}$, which have characteristic durations of \result{$\lesssim 50\,\mathrm{ms}$}.
        (\emph{in order below}) We additionally show $P_\mathrm{joint}$ from the 208 auxiliary channels found in coincidence with the background event as well as $P_\mathrm{min}$ from two individual channels, one found to have a low FAP (\result{L1:ASC-AS\_B\_RF26\_Q\_PIT\_OUT\_DQ, a control channel associated with angular alignment of the interferometer}) and one randomly selected (\result{L1:SUS-SR2\_M3\_MASTER\_OUT\_UL\_DQ, a sensor in the signal-recycling mirror suspension system}).
        Projected histograms show the cumulative distribution of the pointy statistics for each time-series, and grey lines denote the time of the background event, the statistic at that time, and the corresponding FAP.
        We note that, although it was not identified in the search background, there is another, louder whistle approximately two seconds before the background event.
        $P_\mathrm{joint}$ correctly identifies both with \result{$\mathrm{FAP} < 10^{-5}$}.
    }
    \label{fig:whistles}
\end{figure}

Fig.~\ref{fig:whistles} shows $P_\mathrm{joint}$ for a background whistle identified at \result{03:03:15.05 GMT on February 21, 2017} in the LIGO Livingston detector as well as the individual $P_\mathrm{min}$ time-series from significant and insignificant individual auxiliary channels with events found in coincidence.
From the time-series, we see that there is a clear peak in both $P_\mathrm{joint}$ and the significant channel's $P_\mathrm{min}$ at the time of the whistle, but the FAP associated with $P_\mathrm{joint}$ ($\lesssim 10^{-5}$) is lower than the FAP associated with individual $P_\mathrm{min}$ ($\gtrsim 10^{-4}$).
This behavior suggests that each individual witness is polluted by other sources of noise that do not couple to the GW channel in the same way, thereby confounding the inference if only a single channel is used.
However, the additional noise sources are uncorrelated in each auxiliary channel, and the probability of these uncorrelated noise sources occurring simultaneously is low.
Therefore, we can be even more confident that there is a whistle in the GW channel when all the auxiliary channels contain a significant transient.
Additionally, if a single auxiliary channel suddenly stops witnessing whistles, this procedure retains its predictive power, unlike other algorithms that rely on correlations between the GW channel and single auxiliary channels~\cite{Essick2013, Smith2011, Isogai2010}.

This behavior repeats itself for several whistles identified in the search background, typically with similar auxiliary witnesses.
This begs the question of whether a subset of auxiliary witnesses were always coincident with this type of noise, or whether detector non-stationarity led to modified couplings throughout the observing run.
Taking the intersection of each set of \result{$\sim 130$} auxiliary witnesses identified with $P_\mathrm{min} \leq 3\times10^{-2}$ around each whistle produces a much smaller list of only \result{37} channels.
We then produced a na\"{i}ve Bayes $P_\mathrm{joint}$ time-series using only this subset of channels throughout O2 to evaluate its performance.

\begin{figure*}
    \begin{center}
        \includegraphics[width=1.0\textwidth, clip=True, trim=1.05cm 0.15cm 0.60cm 0.75cm]{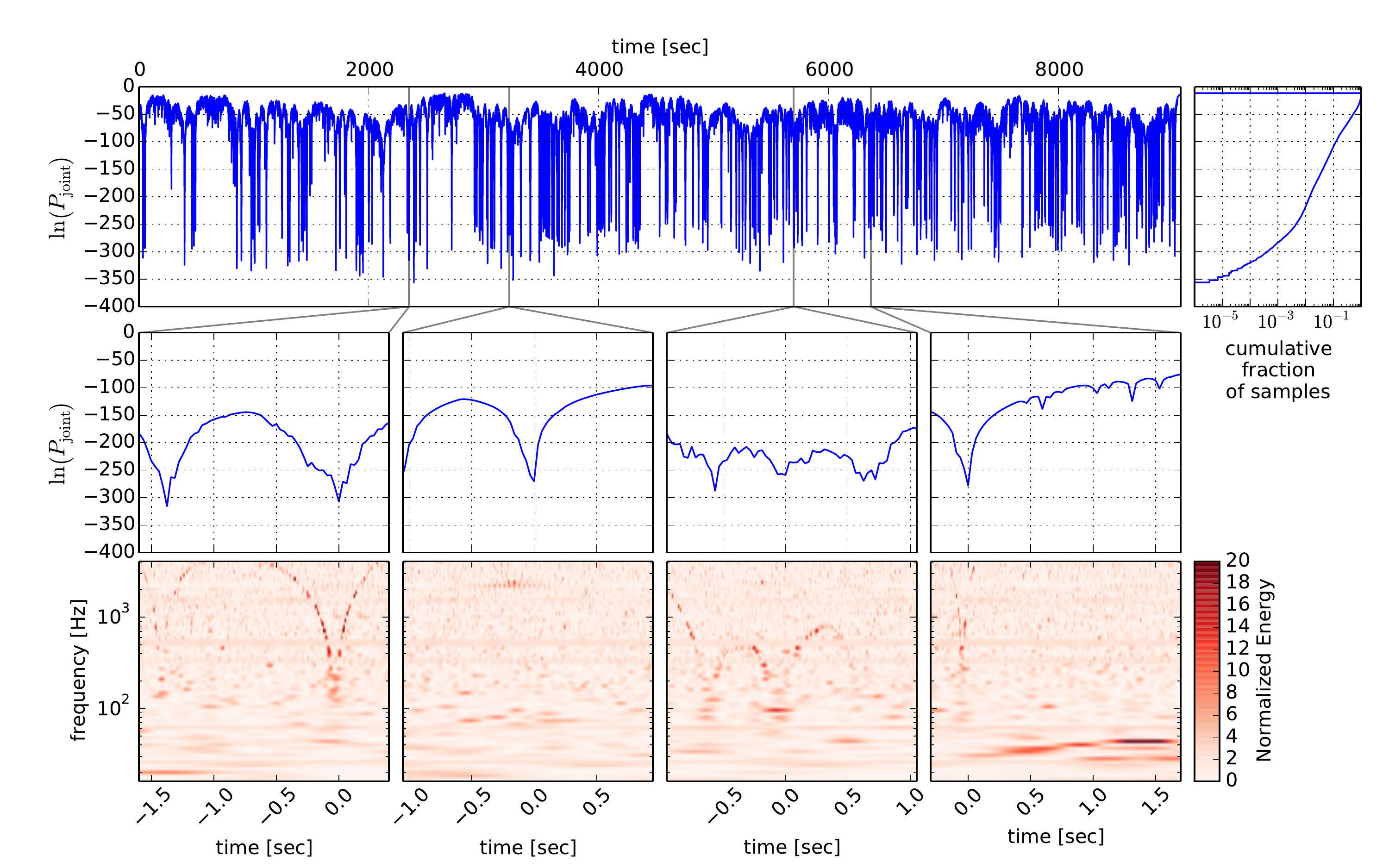}
    \end{center}
    \caption{
        The targeted whistle veto's performance around two of the events identified in the search's background.
        (\emph{top}) $P_\mathrm{joint}$ time-series throughout the $\sim 2.5\,\mathrm{hour}$ stretch.
        Grey shading corresponds to the specific example times called out below.
        (\emph{middle}) Expanded views of $P_\mathrm{joint}$ around four example local minima.
        (\emph{bottom}) Time-frequency representations~\cite{Chatterji2004} of the GW channel at each example.
        We see that a variety of behavior is identified by the veto, including non-stationary coupling to the GW channel: (\emph{left to right}) a pair of whistles with different $\rho$, no apparent non-Gaussian noise, three whistles, and a whistle followed by low-frequency scattering arches (during which there seems to be some activity in a subset of the auxiliary channels).
    }
    \label{fig:whistles too}
\end{figure*}

While the resulting statistic correctly identified the whistles already known from the search background, it showed mixed performance at other times throughout O2.
Indeed, two of the original four events identified in the search background came from the same stretch of data.
Fig.~\ref{fig:whistles too} shows the behavior of $P_\mathrm{joint}$ surrounding those events.
During this \result{$\sim2.5\,\mathrm{hour}$} period between \result{22:21:53 GMT on January 5, 2017 and 00:52:57 GMT on January 6, 2017}, which contains a plethora of non-Gaussian noise, we typically find that the $P_\mathrm{joint}$ derived from this subset of channels clearly identifies whistles at LLO.
However, there appears to be significant variation in the $\rho$ associated with those whistles, somewhat independent of the significance of the $P_\mathrm{joint}$ minima.
This, perhaps, suggests that the auxiliary channels consistently witness a possible source of noise in the GW channel, but the coupling between that noise and the GW channel varies in time, potentially even multiple times per hour (see Fig.~\ref{fig:whistles too}).
Anecdotally, we also found some evidence that the frequency content of noise identified in the GW channel by this set of channels also varied, at times resembling the archetypal whistles shown in Fig.~\ref{fig:whistles} but at other times identifying low-frequency features, not dissimilar to scattering arches \cite{Soni2020}.
This type of non-stationarity implies that, even though a particular time may correspond to an exceptionally rare $P_\mathrm{joint}$, it does not necessarily imply the presence of noise in the GW channel.
As such, the generalization error prevents us from applying this subset of channels to a broader stretch of O2, although the associated FAP could be limited to $\lesssim 10^{-3}$ while identifying a nontrivial fraction of whistles, and we may be forced to identify different sets of witnesses through periodic retraining (c.f.~\cite{Essick2020}).

Nonetheless, we find that the na\"{i}ve Bayes statistic constructed from the intersection of channels identified around background events correctly identifies the two whistles assigned the largest significance within the search's foreground.
Specifically, $P_\mathrm{joint}$ corresponds to \result{$\mathrm{FAP} \lesssim 3\times10^{-4}$} in the neighborhood of each foreground event, even though those events are at least \result{3.5 weeks} away from the nearest of the four background events considered.

Additionally, it is difficult to quantify the fraction of whistles identified by the veto condition due to the difficulty in identifying all whistles present in the detector.
While automatic glitch-classification schemes based on time-frequency representations of the GW channel have been explored (e.g.,~\cite{Zevin2017}), these tools do not provide a simple estimate of the completeness of their catalogs.
That is to say, they do not assign a detection efficiency for identifying individual glitch classes.
Nonetheless, anecdotal evidence suggests that the identified subset of auxiliary channels jointly provide a useful veto condition that identified whistles throughout O2.
Indeed, it does not require deterministic couplings or perfect witnesses, instead relying on the wisdom of a collection of witnesses rather than a single expert.

As a final note, we remark that likelihood ratio tests~\cite{Neyman:1933wgr} may obviate the need to down-select the list of channels in any way.
However, we leave a full exploration of likelihood ratio tests based on the pointy statistic to future work.

\section{Discussion and Conclusions}
\label{sec:discussion}

This work was motivated within the context of GW experiments, in particular the need for a null test to determine the significance of coincidences between GW candidates and auxiliary channels at particular times.
Our statistic can accommodate an arbitrary set of channels, with disparate event rates and amplitudes, as it reduces all measurements to an intuitive probabilistic statement that is directly comparable across all channels.
Even though the formal derivation of our null test assumes events are independent and Poisson-distributed in each channel separately, and this assumption is known to be violated in real-world situations, we nonetheless show our test's utility in several settings.
The power of our null test in establishing the statistical significance of a measurement is retained, and we precisely quantify that significance by directly measuring the background distribution using the same data.

We have shown how our null test can establish which channels in GW detectors can safely be used to construct vetoes by both combining multiple, repeated measurements in a controlled experiment (like hardware injections) and by analyzing \textit{bona fide} astrophysical events.
We find that individual astrophysical events will not be able to overcome the large trials factor from $\mathcal{O}(10^4)$ channels, and typically at least three repeated injections will be needed to limit the expected number of false positives to $\lesssim 1$.
It can also enable targeted studies of transients at specific times in GW detectors, thereby illuminating their couplings and ultimate origin.
We have shown one such case, focusing on non-Gaussian noise artifacts called whistles.
We identify a subset of auxiliary channels that individually correlate with whistles, but together provide a particularly efficient identification scheme during parts of O2.

The utility of such null tests has also proven useful when confirming the astrophysical nature of individual GW candidate events.
Combining the list of auxiliary channels identified as safe with hardware injections with the type of analysis we demonstrated for whistles, one can directly compute the FAP associated with all auxiliary events coincident with a GW candidate.
If all FAPs are large, this constitutes evidence that the behavior of each auxiliary channel is uncorrelated with the GW channel at that time.
Indeed, such analyses were conducted by-hand for the first few GW detections, but further parallelization is needed to scale the pointy analysis to higher detection rates expected from planned detector upgrades~\cite{ObservingScenarios}.

In conclusion, the pointy statistic and methods developed herein are applicable to a wide range of problems beyond GW astrophysics.
Indeed, our null test captures the relevant features of coincidences and naturally quantifies their significance as intuitive probabilistic statements, which are of general use.


\acknowledgments

R.~E. is supported at the University of Chicago by the Kavli Institute for Cosmological Physics through an endowment from the Kavli Foundation and its founder Fred Kavli.
G.~M. and E.~K. are supported by the National Science Foundation (NSF) through award PHY-1764464 to the LIGO Laboratory.
LIGO was constructed by the California Institute of Technology and Massachusetts Institute of Technology with funding from the NSF and operates under cooperative agreement PHY-1764464.
The authors also gratefully acknowledge the computational resources provided by the LIGO Laboratory and supported by NSF grants PHY-0757058 and PHY-0823459.


\bibliography{refs}

\begin{thebibliography}{54}%
\makeatletter
\providecommand \@ifxundefined [1]{%
 \@ifx{#1\undefined}
}%
\providecommand \@ifnum [1]{%
 \ifnum #1\expandafter \@firstoftwo
 \else \expandafter \@secondoftwo
 \fi
}%
\providecommand \@ifx [1]{%
 \ifx #1\expandafter \@firstoftwo
 \else \expandafter \@secondoftwo
 \fi
}%
\providecommand \natexlab [1]{#1}%
\providecommand \enquote  [1]{``#1''}%
\providecommand \bibnamefont  [1]{#1}%
\providecommand \bibfnamefont [1]{#1}%
\providecommand \citenamefont [1]{#1}%
\providecommand \href@noop [0]{\@secondoftwo}%
\providecommand \href [0]{\begingroup \@sanitize@url \@href}%
\providecommand \@href[1]{\@@startlink{#1}\@@href}%
\providecommand \@@href[1]{\endgroup#1\@@endlink}%
\providecommand \@sanitize@url [0]{\catcode `\\12\catcode `\$12\catcode
  `\&12\catcode `\#12\catcode `\^12\catcode `\_12\catcode `\%12\relax}%
\providecommand \@@startlink[1]{}%
\providecommand \@@endlink[0]{}%
\providecommand \url  [0]{\begingroup\@sanitize@url \@url }%
\providecommand \@url [1]{\endgroup\@href {#1}{\urlprefix }}%
\providecommand \urlprefix  [0]{URL }%
\providecommand \Eprint [0]{\href }%
\providecommand \doibase [0]{http://dx.doi.org/}%
\providecommand \selectlanguage [0]{\@gobble}%
\providecommand \bibinfo  [0]{\@secondoftwo}%
\providecommand \bibfield  [0]{\@secondoftwo}%
\providecommand \translation [1]{[#1]}%
\providecommand \BibitemOpen [0]{}%
\providecommand \bibitemStop [0]{}%
\providecommand \bibitemNoStop [0]{.\EOS\space}%
\providecommand \EOS [0]{\spacefactor3000\relax}%
\providecommand \BibitemShut  [1]{\csname bibitem#1\endcsname}%
\let\auto@bib@innerbib\@empty
\bibitem [{\citenamefont {{The Pierre Auger
  Collaboration}}(2015)}]{PierreAuger}%
  \BibitemOpen
  \bibfield  {author} {\bibinfo {author} {\bibnamefont {{The Pierre Auger
  Collaboration}}},\ }\href {\doibase
  https://doi.org/10.1016/j.nima.2015.06.058} {\bibfield  {journal} {\bibinfo
  {journal} {Nuclear Instruments and Methods in Physics Research Section A:
  Accelerators, Spectrometers, Detectors and Associated Equipment}\ }\textbf
  {\bibinfo {volume} {798}},\ \bibinfo {pages} {172 } (\bibinfo {year}
  {2015})}\BibitemShut {NoStop}%
\bibitem [{\citenamefont {DeYoung}(2012)}]{HAWC}%
  \BibitemOpen
  \bibfield  {author} {\bibinfo {author} {\bibfnamefont {T.}~\bibnamefont
  {DeYoung}},\ }\href {\doibase https://doi.org/10.1016/j.nima.2012.01.026}
  {\bibfield  {journal} {\bibinfo  {journal} {Nuclear Instruments and Methods
  in Physics Research Section A: Accelerators, Spectrometers, Detectors and
  Associated Equipment}\ }\textbf {\bibinfo {volume} {692}},\ \bibinfo {pages}
  {72 } (\bibinfo {year} {2012})},\ \bibinfo {note} {3rd Roma International
  Conference on Astroparticle Physics}\BibitemShut {NoStop}%
\bibitem [{\citenamefont {Halzen}\ and\ \citenamefont {Klein}(2010)}]{IceCube}%
  \BibitemOpen
  \bibfield  {author} {\bibinfo {author} {\bibfnamefont {F.}~\bibnamefont
  {Halzen}}\ and\ \bibinfo {author} {\bibfnamefont {S.~R.}\ \bibnamefont
  {Klein}},\ }\href {\doibase 10.1063/1.3480478} {\bibfield  {journal}
  {\bibinfo  {journal} {Review of Scientific Instruments}\ }\textbf {\bibinfo
  {volume} {81}},\ \bibinfo {pages} {081101} (\bibinfo {year} {2010})},\
  \Eprint {http://arxiv.org/abs/https://doi.org/10.1063/1.3480478}
  {https://doi.org/10.1063/1.3480478} \BibitemShut {NoStop}%
\bibitem [{\citenamefont {Ageron}\ \emph {et~al.}(2011)\citenamefont {Ageron}
  \emph {et~al.}}]{Antares}%
  \BibitemOpen
  \bibfield  {author} {\bibinfo {author} {\bibfnamefont {M.}~\bibnamefont
  {Ageron}} \emph {et~al.},\ }\href {\doibase
  https://doi.org/10.1016/j.nima.2011.06.103} {\bibfield  {journal} {\bibinfo
  {journal} {Nuclear Instruments and Methods in Physics Research Section A:
  Accelerators, Spectrometers, Detectors and Associated Equipment}\ }\textbf
  {\bibinfo {volume} {656}},\ \bibinfo {pages} {11 } (\bibinfo {year}
  {2011})}\BibitemShut {NoStop}%
\bibitem [{\citenamefont {Antonioli}\ \emph {et~al.}(2004)\citenamefont
  {Antonioli} \emph {et~al.}}]{SNEWS}%
  \BibitemOpen
  \bibfield  {author} {\bibinfo {author} {\bibfnamefont {P.}~\bibnamefont
  {Antonioli}} \emph {et~al.},\ }\href {\doibase 10.1088/1367-2630/6/1/114}
  {\bibfield  {journal} {\bibinfo  {journal} {New Journal of Physics}\ }\textbf
  {\bibinfo {volume} {6}},\ \bibinfo {pages} {114} (\bibinfo {year}
  {2004})}\BibitemShut {NoStop}%
\bibitem [{\citenamefont {Schumann}(1952{\natexlab{a}})}]{Schumann1}%
  \BibitemOpen
  \bibfield  {author} {\bibinfo {author} {\bibfnamefont {W.~O.}\ \bibnamefont
  {Schumann}},\ }\href
  {https://www.degruyter.com/view/journals/zna/7/2/article-p149.xml} {\bibfield
   {journal} {\bibinfo  {journal} {Zeitschrift für Naturforschung A}\ }\textbf
  {\bibinfo {volume} {7}},\ \bibinfo {pages} {149 } (\bibinfo {year}
  {1952}{\natexlab{a}})}\BibitemShut {NoStop}%
\bibitem [{\citenamefont {Schumann}(1952{\natexlab{b}})}]{Schumann2}%
  \BibitemOpen
  \bibfield  {author} {\bibinfo {author} {\bibfnamefont {W.~O.}\ \bibnamefont
  {Schumann}},\ }\href
  {https://www.degruyter.com/view/journals/zna/7/3-4/article-p250.xml}
  {\bibfield  {journal} {\bibinfo  {journal} {Zeitschrift für Naturforschung
  A}\ }\textbf {\bibinfo {volume} {7}},\ \bibinfo {pages} {250 } (\bibinfo
  {year} {1952}{\natexlab{b}})}\BibitemShut {NoStop}%
\bibitem [{\citenamefont {Thrane}\ \emph {et~al.}(2013)\citenamefont {Thrane},
  \citenamefont {Christensen},\ and\ \citenamefont {Schofield}}]{Thrane2013}%
  \BibitemOpen
  \bibfield  {author} {\bibinfo {author} {\bibfnamefont {E.}~\bibnamefont
  {Thrane}}, \bibinfo {author} {\bibfnamefont {N.}~\bibnamefont {Christensen}},
  \ and\ \bibinfo {author} {\bibfnamefont {R.~M.~S.}\ \bibnamefont
  {Schofield}},\ }\href {\doibase 10.1103/PhysRevD.87.123009} {\bibfield
  {journal} {\bibinfo  {journal} {Phys. Rev. D}\ }\textbf {\bibinfo {volume}
  {87}},\ \bibinfo {pages} {123009} (\bibinfo {year} {2013})}\BibitemShut
  {NoStop}%
\bibitem [{\citenamefont {Abbott}\ \emph
  {et~al.}(2017{\natexlab{a}})\citenamefont {Abbott} \emph
  {et~al.}}]{GW170817MMA}%
  \BibitemOpen
  \bibfield  {author} {\bibinfo {author} {\bibfnamefont {B.~P.}\ \bibnamefont
  {Abbott}} \emph {et~al.},\ }\href {\doibase 10.3847/2041-8213/aa91c9}
  {\bibfield  {journal} {\bibinfo  {journal} {The Astrophysical Journal}\
  }\textbf {\bibinfo {volume} {848}},\ \bibinfo {pages} {L12} (\bibinfo {year}
  {2017}{\natexlab{a}})}\BibitemShut {NoStop}%
\bibitem [{\citenamefont {Biswas}\ \emph {et~al.}(2013)\citenamefont {Biswas}
  \emph {et~al.}}]{Biswas2013}%
  \BibitemOpen
  \bibfield  {author} {\bibinfo {author} {\bibfnamefont {R.}~\bibnamefont
  {Biswas}} \emph {et~al.},\ }\href {\doibase 10.1103/PhysRevD.88.062003}
  {\bibfield  {journal} {\bibinfo  {journal} {Phys. Rev. D}\ }\textbf {\bibinfo
  {volume} {88}},\ \bibinfo {pages} {062003} (\bibinfo {year}
  {2013})}\BibitemShut {NoStop}%
\bibitem [{\citenamefont {Essick}\ \emph {et~al.}(2013)\citenamefont {Essick},
  \citenamefont {Blackburn},\ and\ \citenamefont {Katsavounidis}}]{Essick2013}%
  \BibitemOpen
  \bibfield  {author} {\bibinfo {author} {\bibfnamefont {R.}~\bibnamefont
  {Essick}}, \bibinfo {author} {\bibfnamefont {L.}~\bibnamefont {Blackburn}}, \
  and\ \bibinfo {author} {\bibfnamefont {E.}~\bibnamefont {Katsavounidis}},\
  }\href {\doibase 10.1088/0264-9381/30/15/155010} {\bibfield  {journal}
  {\bibinfo  {journal} {Classical and Quantum Gravity}\ }\textbf {\bibinfo
  {volume} {30}},\ \bibinfo {pages} {155010} (\bibinfo {year}
  {2013})}\BibitemShut {NoStop}%
\bibitem [{\citenamefont {{Essick}}\ \emph {et~al.}(2020)\citenamefont
  {{Essick}} \emph {et~al.}}]{Essick2020}%
  \BibitemOpen
  \bibfield  {author} {\bibinfo {author} {\bibfnamefont {R.}~\bibnamefont
  {{Essick}}} \emph {et~al.},\ }\href@noop {} {\bibfield  {journal} {\bibinfo
  {journal} {arXiv e-prints}\ ,\ \bibinfo {eid} {arXiv:2005.12761}} (\bibinfo
  {year} {2020})},\ \Eprint {http://arxiv.org/abs/2005.12761} {arXiv:2005.12761
  [astro-ph.IM]} \BibitemShut {NoStop}%
\bibitem [{\citenamefont {Bhandari}\ \emph {et~al.}(2016)\citenamefont
  {Bhandari}, \citenamefont {Sangal},\ and\ \citenamefont
  {Kumar}}]{Bhandari2016}%
  \BibitemOpen
  \bibfield  {author} {\bibinfo {author} {\bibfnamefont {A.}~\bibnamefont
  {Bhandari}}, \bibinfo {author} {\bibfnamefont {A.~L.}\ \bibnamefont
  {Sangal}}, \ and\ \bibinfo {author} {\bibfnamefont {K.}~\bibnamefont
  {Kumar}},\ }\href {\doibase 10.1002/sec.1472} {\bibfield  {journal} {\bibinfo
   {journal} {Security and Communication Networks}\ }\textbf {\bibinfo {volume}
  {9}},\ \bibinfo {pages} {2222} (\bibinfo {year} {2016})},\ \Eprint
  {http://arxiv.org/abs/https://onlinelibrary.wiley.com/doi/pdf/10.1002/sec.1472}
  {https://onlinelibrary.wiley.com/doi/pdf/10.1002/sec.1472} \BibitemShut
  {NoStop}%
\bibitem [{\citenamefont {{Bouchaud}}(2009)}]{Bouchaud2009}%
  \BibitemOpen
  \bibfield  {author} {\bibinfo {author} {\bibfnamefont {J.~P.}\ \bibnamefont
  {{Bouchaud}}},\ }\href@noop {} {\bibfield  {journal} {\bibinfo  {journal}
  {arXiv e-prints}\ ,\ \bibinfo {eid} {arXiv:0903.2428}} (\bibinfo {year}
  {2009})},\ \Eprint {http://arxiv.org/abs/0903.2428} {arXiv:0903.2428
  [q-fin.TR]} \BibitemShut {NoStop}%
\bibitem [{\citenamefont {Poon}\ \emph {et~al.}(2004)\citenamefont {Poon},
  \citenamefont {Rockinger},\ and\ \citenamefont {Tawn}}]{Poon2004}%
  \BibitemOpen
  \bibfield  {author} {\bibinfo {author} {\bibfnamefont {S.}~\bibnamefont
  {Poon}}, \bibinfo {author} {\bibfnamefont {M.}~\bibnamefont {Rockinger}}, \
  and\ \bibinfo {author} {\bibfnamefont {J.}~\bibnamefont {Tawn}},\ }\href@noop
  {} {\bibfield  {journal} {\bibinfo  {journal} {Review of Financial Studies}\
  }\textbf {\bibinfo {volume} {17}},\ \bibinfo {pages} {581} (\bibinfo {year}
  {2004})}\BibitemShut {NoStop}%
\bibitem [{\citenamefont {Aasi}\ \emph {et~al.}(2015)\citenamefont {Aasi} \emph
  {et~al.}}]{LIGO}%
  \BibitemOpen
  \bibfield  {author} {\bibinfo {author} {\bibfnamefont {J.}~\bibnamefont
  {Aasi}} \emph {et~al.},\ }\href {\doibase 10.1088/0264-9381/32/7/074001}
  {\bibfield  {journal} {\bibinfo  {journal} {Class. Quantum Grav.}\ }\textbf
  {\bibinfo {volume} {32}},\ \bibinfo {eid} {074001} (\bibinfo {year}
  {2015})},\ \Eprint {http://arxiv.org/abs/1411.4547} {arXiv:1411.4547 [gr-qc]}
  \BibitemShut {NoStop}%
\bibitem [{\citenamefont {{F. Acernese \textit{et al.} (Virgo
  Collaboration)}}(2015)}]{Virgo}%
  \BibitemOpen
  \bibfield  {author} {\bibinfo {author} {\bibnamefont {{F. Acernese \textit{et
  al.} (Virgo Collaboration)}}},\ }\href {\doibase
  10.1088/0264-9381/32/2/024001} {\bibfield  {journal} {\bibinfo  {journal}
  {Class. Quantum Grav.}\ }\textbf {\bibinfo {volume} {32}},\ \bibinfo {eid}
  {024001} (\bibinfo {year} {2015})},\ \Eprint {http://arxiv.org/abs/1408.3978}
  {arXiv:1408.3978 [gr-qc]} \BibitemShut {NoStop}%
\bibitem [{\citenamefont {Isi}\ \emph {et~al.}(2018)\citenamefont {Isi} \emph
  {et~al.}}]{Isi2018}%
  \BibitemOpen
  \bibfield  {author} {\bibinfo {author} {\bibfnamefont {M.}~\bibnamefont
  {Isi}} \emph {et~al.},\ }\href {\doibase 10.1103/PhysRevD.98.042007}
  {\bibfield  {journal} {\bibinfo  {journal} {Phys. Rev. D}\ }\textbf {\bibinfo
  {volume} {98}},\ \bibinfo {pages} {042007} (\bibinfo {year}
  {2018})}\BibitemShut {NoStop}%
\bibitem [{\citenamefont {Cornish}\ and\ \citenamefont
  {Littenberg}(2015)}]{Cornish2015}%
  \BibitemOpen
  \bibfield  {author} {\bibinfo {author} {\bibfnamefont {N.~J.}\ \bibnamefont
  {Cornish}}\ and\ \bibinfo {author} {\bibfnamefont {T.~B.}\ \bibnamefont
  {Littenberg}},\ }\href {\doibase 10.1088/0264-9381/32/13/135012} {\bibfield
  {journal} {\bibinfo  {journal} {Classical and Quantum Gravity}\ }\textbf
  {\bibinfo {volume} {32}},\ \bibinfo {pages} {135012} (\bibinfo {year}
  {2015})}\BibitemShut {NoStop}%
\bibitem [{\citenamefont {Kanner}\ \emph {et~al.}(2016)\citenamefont {Kanner}
  \emph {et~al.}}]{Kanner2016}%
  \BibitemOpen
  \bibfield  {author} {\bibinfo {author} {\bibfnamefont {J.~B.}\ \bibnamefont
  {Kanner}} \emph {et~al.},\ }\href {\doibase 10.1103/PhysRevD.93.022002}
  {\bibfield  {journal} {\bibinfo  {journal} {Phys. Rev. D}\ }\textbf {\bibinfo
  {volume} {93}},\ \bibinfo {pages} {022002} (\bibinfo {year}
  {2016})}\BibitemShut {NoStop}%
\bibitem [{\citenamefont {Buonanno}\ \emph {et~al.}(2009)\citenamefont
  {Buonanno} \emph {et~al.}}]{Buonanno2009}%
  \BibitemOpen
  \bibfield  {author} {\bibinfo {author} {\bibfnamefont {A.}~\bibnamefont
  {Buonanno}} \emph {et~al.},\ }\href {\doibase 10.1103/PhysRevD.80.084043}
  {\bibfield  {journal} {\bibinfo  {journal} {Phys. Rev. D}\ }\textbf {\bibinfo
  {volume} {80}},\ \bibinfo {pages} {084043} (\bibinfo {year}
  {2009})}\BibitemShut {NoStop}%
\bibitem [{\citenamefont {Cahillane}\ \emph {et~al.}(2017)\citenamefont
  {Cahillane} \emph {et~al.}}]{Cahillane2017}%
  \BibitemOpen
  \bibfield  {author} {\bibinfo {author} {\bibfnamefont {C.}~\bibnamefont
  {Cahillane}} \emph {et~al.},\ }\href {\doibase 10.1103/PhysRevD.96.102001}
  {\bibfield  {journal} {\bibinfo  {journal} {Phys. Rev. D}\ }\textbf {\bibinfo
  {volume} {96}},\ \bibinfo {pages} {102001} (\bibinfo {year}
  {2017})}\BibitemShut {NoStop}%
\bibitem [{\citenamefont {Essick}\ and\ \citenamefont
  {Holz}(2019)}]{Essick2019}%
  \BibitemOpen
  \bibfield  {author} {\bibinfo {author} {\bibfnamefont {R.}~\bibnamefont
  {Essick}}\ and\ \bibinfo {author} {\bibfnamefont {D.~E.}\ \bibnamefont
  {Holz}},\ }\href {\doibase 10.1088/1361-6382/ab2142} {\bibfield  {journal}
  {\bibinfo  {journal} {Classical and Quantum Gravity}\ }\textbf {\bibinfo
  {volume} {36}},\ \bibinfo {pages} {125002} (\bibinfo {year}
  {2019})}\BibitemShut {NoStop}%
\bibitem [{\citenamefont {{Abbott}}\ \emph {et~al.}(2018)\citenamefont
  {{Abbott}} \emph {et~al.}}]{GWTC-1}%
  \BibitemOpen
  \bibfield  {author} {\bibinfo {author} {\bibfnamefont {B.~P.}\ \bibnamefont
  {{Abbott}}} \emph {et~al.},\ }\href@noop {} {\bibfield  {journal} {\bibinfo
  {journal} {arXiv e-prints}\ ,\ \bibinfo {eid} {arXiv:1811.12907}} (\bibinfo
  {year} {2018})},\ \Eprint {http://arxiv.org/abs/1811.12907} {arXiv:1811.12907
  [astro-ph.HE]} \BibitemShut {NoStop}%
\bibitem [{\citenamefont {Abbott}\ \emph {et~al.}(2020)\citenamefont {Abbott}
  \emph {et~al.}}]{GWTC-2}%
  \BibitemOpen
  \bibfield  {author} {\bibinfo {author} {\bibfnamefont {R.}~\bibnamefont
  {Abbott}} \emph {et~al.},\ }\href@noop {} {\  (\bibinfo {year} {2020})},\
  \Eprint {http://arxiv.org/abs/2010.14527} {arXiv:2010.14527 [gr-qc]}
  \BibitemShut {NoStop}%
\bibitem [{\citenamefont {Chatterji}\ \emph {et~al.}(2004)\citenamefont
  {Chatterji} \emph {et~al.}}]{Chatterji2004}%
  \BibitemOpen
  \bibfield  {author} {\bibinfo {author} {\bibfnamefont {S.}~\bibnamefont
  {Chatterji}} \emph {et~al.},\ }\href {\doibase 10.1088/0264-9381/21/20/024}
  {\bibfield  {journal} {\bibinfo  {journal} {Classical and Quantum Gravity}\
  }\textbf {\bibinfo {volume} {21}},\ \bibinfo {pages} {S1809} (\bibinfo {year}
  {2004})}\BibitemShut {NoStop}%
\bibitem [{\citenamefont {Godwin}(2020)}]{godwin-thesis}%
  \BibitemOpen
  \bibfield  {author} {\bibinfo {author} {\bibfnamefont {P.}~\bibnamefont
  {Godwin}},\ }\href@noop {} {\enquote {\bibinfo {title} {Low-latency
  statistical data quality in the era of multi-messenger astronomy},}\ }
  (\bibinfo {year} {2020})\BibitemShut {NoStop}%
\bibitem [{\citenamefont {Sun}\ \emph {et~al.}(2020)\citenamefont {Sun} \emph
  {et~al.}}]{Sun_2020}%
  \BibitemOpen
  \bibfield  {author} {\bibinfo {author} {\bibfnamefont {L.}~\bibnamefont
  {Sun}} \emph {et~al.},\ }\href {\doibase 10.1088/1361-6382/abb14e} {\bibfield
   {journal} {\bibinfo  {journal} {Classical and Quantum Gravity}\ }\textbf
  {\bibinfo {volume} {37}},\ \bibinfo {pages} {225008} (\bibinfo {year}
  {2020})}\BibitemShut {NoStop}%
\bibitem [{\citenamefont {Abbott}\ \emph {et~al.}(2018)\citenamefont {Abbott}
  \emph {et~al.}}]{Abbott2018}%
  \BibitemOpen
  \bibfield  {author} {\bibinfo {author} {\bibfnamefont {B.~P.}\ \bibnamefont
  {Abbott}} \emph {et~al.},\ }\href {\doibase 10.1088/1361-6382/aaaafa}
  {\bibfield  {journal} {\bibinfo  {journal} {Classical and Quantum Gravity}\
  }\textbf {\bibinfo {volume} {35}},\ \bibinfo {pages} {065010} (\bibinfo
  {year} {2018})}\BibitemShut {NoStop}%
\bibitem [{\citenamefont {Cabero}\ \emph {et~al.}(2019)\citenamefont {Cabero}
  \emph {et~al.}}]{Cabero2019}%
  \BibitemOpen
  \bibfield  {author} {\bibinfo {author} {\bibfnamefont {M.}~\bibnamefont
  {Cabero}} \emph {et~al.},\ }\href {\doibase 10.1088/1361-6382/ab2e14}
  {\bibfield  {journal} {\bibinfo  {journal} {Classical and Quantum Gravity}\
  }\textbf {\bibinfo {volume} {36}},\ \bibinfo {pages} {155010} (\bibinfo
  {year} {2019})}\BibitemShut {NoStop}%
\bibitem [{\citenamefont {Zevin}\ \emph {et~al.}(2017)\citenamefont {Zevin}
  \emph {et~al.}}]{Zevin2017}%
  \BibitemOpen
  \bibfield  {author} {\bibinfo {author} {\bibfnamefont {M.}~\bibnamefont
  {Zevin}} \emph {et~al.},\ }\href {\doibase 10.1088/1361-6382/aa5cea}
  {\bibfield  {journal} {\bibinfo  {journal} {Classical and Quantum Gravity}\
  }\textbf {\bibinfo {volume} {34}},\ \bibinfo {pages} {064003} (\bibinfo
  {year} {2017})}\BibitemShut {NoStop}%
\bibitem [{\citenamefont {Effler}\ \emph {et~al.}(2015)\citenamefont {Effler}
  \emph {et~al.}}]{Effler:2014zpa}%
  \BibitemOpen
  \bibfield  {author} {\bibinfo {author} {\bibfnamefont {A.}~\bibnamefont
  {Effler}} \emph {et~al.},\ }\href {\doibase 10.1088/0264-9381/32/3/035017}
  {\bibfield  {journal} {\bibinfo  {journal} {Class. Quant. Grav.}\ }\textbf
  {\bibinfo {volume} {32}},\ \bibinfo {pages} {035017} (\bibinfo {year}
  {2015})},\ \Eprint {http://arxiv.org/abs/1409.5160} {arXiv:1409.5160
  [astro-ph.IM]} \BibitemShut {NoStop}%
\bibitem [{\citenamefont {Smith}\ \emph {et~al.}(2011)\citenamefont {Smith}
  \emph {et~al.}}]{Smith2011}%
  \BibitemOpen
  \bibfield  {author} {\bibinfo {author} {\bibfnamefont {J.~R.}\ \bibnamefont
  {Smith}} \emph {et~al.},\ }\href {\doibase 10.1088/0264-9381/28/23/235005}
  {\bibfield  {journal} {\bibinfo  {journal} {Classical and Quantum Gravity}\
  }\textbf {\bibinfo {volume} {28}},\ \bibinfo {pages} {235005} (\bibinfo
  {year} {2011})}\BibitemShut {NoStop}%
\bibitem [{\citenamefont {Isogai}\ \emph {et~al.}(2010)\citenamefont {Isogai}
  \emph {et~al.}}]{Isogai2010}%
  \BibitemOpen
  \bibfield  {author} {\bibinfo {author} {\bibfnamefont {T.}~\bibnamefont
  {Isogai}} \emph {et~al.},\ }\href {\doibase 10.1088/1742-6596/243/1/012005}
  {\bibfield  {journal} {\bibinfo  {journal} {Journal of Physics: Conference
  Series}\ }\textbf {\bibinfo {volume} {243}},\ \bibinfo {pages} {012005}
  (\bibinfo {year} {2010})}\BibitemShut {NoStop}%
\bibitem [{\citenamefont {{Cuoco}}\ \emph {et~al.}(2020)\citenamefont {{Cuoco}}
  \emph {et~al.}}]{Cuoco2020}%
  \BibitemOpen
  \bibfield  {author} {\bibinfo {author} {\bibfnamefont {E.}~\bibnamefont
  {{Cuoco}}} \emph {et~al.},\ }\href@noop {} {\bibfield  {journal} {\bibinfo
  {journal} {arXiv e-prints}\ ,\ \bibinfo {eid} {arXiv:2005.03745}} (\bibinfo
  {year} {2020})},\ \Eprint {http://arxiv.org/abs/2005.03745} {arXiv:2005.03745
  [astro-ph.HE]} \BibitemShut {NoStop}%
\bibitem [{\citenamefont {Connaughton}\ \emph {et~al.}(2016)\citenamefont
  {Connaughton} \emph {et~al.}}]{Connaughton2016}%
  \BibitemOpen
  \bibfield  {author} {\bibinfo {author} {\bibfnamefont {V.}~\bibnamefont
  {Connaughton}} \emph {et~al.},\ }\href {\doibase 10.3847/2041-8205/826/1/l6}
  {\bibfield  {journal} {\bibinfo  {journal} {The Astrophysical Journal}\
  }\textbf {\bibinfo {volume} {826}},\ \bibinfo {pages} {L6} (\bibinfo {year}
  {2016})}\BibitemShut {NoStop}%
\bibitem [{\citenamefont {{Lynch}}\ \emph {et~al.}(2018)\citenamefont
  {{Lynch}}, \citenamefont {{Vitale}},\ and\ \citenamefont
  {{Katsavounidis}}}]{Lynch2018}%
  \BibitemOpen
  \bibfield  {author} {\bibinfo {author} {\bibfnamefont {R.}~\bibnamefont
  {{Lynch}}}, \bibinfo {author} {\bibfnamefont {S.}~\bibnamefont {{Vitale}}}, \
  and\ \bibinfo {author} {\bibfnamefont {E.}~\bibnamefont {{Katsavounidis}}},\
  }\href@noop {} {\bibfield  {journal} {\bibinfo  {journal} {arXiv e-prints}\
  ,\ \bibinfo {eid} {arXiv:1811.01297}} (\bibinfo {year} {2018})},\ \Eprint
  {http://arxiv.org/abs/1811.01297} {arXiv:1811.01297 [physics.data-an]}
  \BibitemShut {NoStop}%
\bibitem [{\citenamefont {Mueller}\ \emph {et~al.}(2016)\citenamefont {Mueller}
  \emph {et~al.}}]{Mueller:2016hex}%
  \BibitemOpen
  \bibfield  {author} {\bibinfo {author} {\bibfnamefont {C.~L.}\ \bibnamefont
  {Mueller}} \emph {et~al.} (\bibinfo {collaboration} {aLIGO}),\ }\href
  {\doibase 10.1063/1.4936974} {\bibfield  {journal} {\bibinfo  {journal} {Rev.
  Sci. Instrum.}\ }\textbf {\bibinfo {volume} {87}},\ \bibinfo {pages} {014502}
  (\bibinfo {year} {2016})},\ \Eprint {http://arxiv.org/abs/1601.05442}
  {arXiv:1601.05442 [physics.ins-det]} \BibitemShut {NoStop}%
\bibitem [{\citenamefont {Matichard}\ \emph {et~al.}(2015)\citenamefont
  {Matichard} \emph {et~al.}}]{Matichard:2015eva}%
  \BibitemOpen
  \bibfield  {author} {\bibinfo {author} {\bibfnamefont {F.}~\bibnamefont
  {Matichard}} \emph {et~al.},\ }\href {\doibase
  10.1088/0264-9381/32/18/185003} {\bibfield  {journal} {\bibinfo  {journal}
  {Class. Quant. Grav.}\ }\textbf {\bibinfo {volume} {32}},\ \bibinfo {pages}
  {185003} (\bibinfo {year} {2015})},\ \Eprint
  {http://arxiv.org/abs/1502.06300} {arXiv:1502.06300 [physics.ins-det]}
  \BibitemShut {NoStop}%
\bibitem [{\citenamefont {Staley}(2015)}]{Staley:2015nie}%
  \BibitemOpen
  \bibfield  {author} {\bibinfo {author} {\bibfnamefont {A.}~\bibnamefont
  {Staley}},\ }\emph {\bibinfo {title} {{Locking the Advanced LIGO
  Gravitational Wave Detector: with a focus on the Arm Length Stabilization
  Technique}}},\ \href {\doibase 10.7916/D8X34WQ4} {Ph.D. thesis},\ \bibinfo
  {school} {Columbia U.} (\bibinfo {year} {2015})\BibitemShut {NoStop}%
\bibitem [{\citenamefont {Graef~Rollins}(2016)}]{Rollins:2016hlk}%
  \BibitemOpen
  \bibfield  {author} {\bibinfo {author} {\bibfnamefont {J.}~\bibnamefont
  {Graef~Rollins}},\ }\href@noop {} {\  (\bibinfo {year} {2016})},\ \Eprint
  {http://arxiv.org/abs/1604.01456} {arXiv:1604.01456 [astro-ph.IM]}
  \BibitemShut {NoStop}%
\bibitem [{\citenamefont {{Zackay}}\ \emph {et~al.}(2019)\citenamefont
  {{Zackay}} \emph {et~al.}}]{Zackay:2019}%
  \BibitemOpen
  \bibfield  {author} {\bibinfo {author} {\bibfnamefont {B.}~\bibnamefont
  {{Zackay}}} \emph {et~al.},\ }\href@noop {} {\bibfield  {journal} {\bibinfo
  {journal} {arXiv e-prints}\ ,\ \bibinfo {eid} {arXiv:1908.05644}} (\bibinfo
  {year} {2019})},\ \Eprint {http://arxiv.org/abs/1908.05644} {arXiv:1908.05644
  [astro-ph.IM]} \BibitemShut {NoStop}%
\bibitem [{\citenamefont {Essick}(2019)}]{LLOalog}%
  \BibitemOpen
  \bibfield  {author} {\bibinfo {author} {\bibfnamefont {R.}~\bibnamefont
  {Essick}},\ }\href@noop {} {\enquote {\bibinfo {title} {Detchar safety
  hardware injections starting at 21:00:00 utc 3 sep 2019},}\ }\bibinfo
  {howpublished}
  {\url{https://alog.ligo-la.caltech.edu/aLOG/index.php?callRep=48277}}
  (\bibinfo {year} {2019})\BibitemShut {NoStop}%
\bibitem [{\citenamefont {Blackburn}(2007)}]{kleine-welle}%
  \BibitemOpen
  \bibfield  {author} {\bibinfo {author} {\bibfnamefont {L.}~\bibnamefont
  {Blackburn}},\ }\href@noop {} {\enquote {\bibinfo {title} {Kleine-welle
  algorithm},}\ }\bibinfo {howpublished}
  {\url{https://dcc.ligo.org/public/0027/T060221/000/T060221-00.pdf}} (\bibinfo
  {year} {2007})\BibitemShut {NoStop}%
\bibitem [{\citenamefont {Abbott}\ \emph
  {et~al.}(2017{\natexlab{b}})\citenamefont {Abbott} \emph
  {et~al.}}]{GW170814}%
  \BibitemOpen
  \bibfield  {author} {\bibinfo {author} {\bibfnamefont {B.~P.}\ \bibnamefont
  {Abbott}} \emph {et~al.},\ }\href {\doibase 10.1103/PhysRevLett.119.141101}
  {\bibfield  {journal} {\bibinfo  {journal} {Phys. Rev. Lett.}\ }\textbf
  {\bibinfo {volume} {119}},\ \bibinfo {pages} {141101} (\bibinfo {year}
  {2017}{\natexlab{b}})}\BibitemShut {NoStop}%
\bibitem [{\citenamefont {{The LIGO Scientific
  Collaboration}}(2020)}]{GW170814-aux-data}%
  \BibitemOpen
  \bibfield  {author} {\bibinfo {author} {\bibnamefont {{The LIGO Scientific
  Collaboration}}},\ }\href@noop {} {\enquote {\bibinfo {title} {Auxiliary
  channel three hour release},}\ }\bibinfo {howpublished}
  {https://www.gw-openscience.org/auxiliary/GW170814/} (\bibinfo {year}
  {2020})\BibitemShut {NoStop}%
\bibitem [{\citenamefont {Lynch}\ \emph {et~al.}(2017)\citenamefont {Lynch}
  \emph {et~al.}}]{Lynch2017}%
  \BibitemOpen
  \bibfield  {author} {\bibinfo {author} {\bibfnamefont {R.}~\bibnamefont
  {Lynch}} \emph {et~al.},\ }\href {\doibase 10.1103/PhysRevD.95.104046}
  {\bibfield  {journal} {\bibinfo  {journal} {Phys. Rev. D}\ }\textbf {\bibinfo
  {volume} {95}},\ \bibinfo {pages} {104046} (\bibinfo {year}
  {2017})}\BibitemShut {NoStop}%
\bibitem [{\citenamefont {Abbott}\ \emph {et~al.}(2019)\citenamefont {Abbott}
  \emph {et~al.}}]{O2burst}%
  \BibitemOpen
  \bibfield  {author} {\bibinfo {author} {\bibfnamefont {B.~P.}\ \bibnamefont
  {Abbott}} \emph {et~al.},\ }\href {\doibase 10.1103/PhysRevD.100.024017}
  {\bibfield  {journal} {\bibinfo  {journal} {Phys. Rev. D}\ }\textbf {\bibinfo
  {volume} {100}},\ \bibinfo {pages} {024017} (\bibinfo {year}
  {2019})}\BibitemShut {NoStop}%
\bibitem [{\citenamefont {{Soni}}\ \emph {et~al.}(2020)\citenamefont {{Soni}}
  \emph {et~al.}}]{Soni2020}%
  \BibitemOpen
  \bibfield  {author} {\bibinfo {author} {\bibfnamefont {S.}~\bibnamefont
  {{Soni}}} \emph {et~al.},\ }\href@noop {} {\bibfield  {journal} {\bibinfo
  {journal} {arXiv e-prints}\ ,\ \bibinfo {eid} {arXiv:2007.14876}} (\bibinfo
  {year} {2020})},\ \Eprint {http://arxiv.org/abs/2007.14876} {arXiv:2007.14876
  [astro-ph.IM]} \BibitemShut {NoStop}%
\bibitem [{\citenamefont {Neyman}\ and\ \citenamefont
  {Pearson}(1933)}]{Neyman:1933wgr}%
  \BibitemOpen
  \bibfield  {author} {\bibinfo {author} {\bibfnamefont {J.}~\bibnamefont
  {Neyman}}\ and\ \bibinfo {author} {\bibfnamefont {E.~S.}\ \bibnamefont
  {Pearson}},\ }\href {\doibase 10.1098/rsta.1933.0009} {\bibfield  {journal}
  {\bibinfo  {journal} {Phil. Trans. Roy. Soc. Lond.}\ }\textbf {\bibinfo
  {volume} {A231}},\ \bibinfo {pages} {289} (\bibinfo {year}
  {1933})}\BibitemShut {NoStop}%
\bibitem [{\citenamefont {{Abbott}}\ \emph {et~al.}(2018)\citenamefont
  {{Abbott}} \emph {et~al.}}]{ObservingScenarios}%
  \BibitemOpen
  \bibfield  {author} {\bibinfo {author} {\bibfnamefont {B.~P.}\ \bibnamefont
  {{Abbott}}} \emph {et~al.},\ }\href {\doibase 10.1007/s41114-018-0012-9}
  {\bibfield  {journal} {\bibinfo  {journal} {Living Reviews in Relativity}\
  }\textbf {\bibinfo {volume} {21}},\ \bibinfo {eid} {3} (\bibinfo {year}
  {2018})},\ \Eprint {http://arxiv.org/abs/1304.0670} {arXiv:1304.0670 [gr-qc]}
  \BibitemShut {NoStop}%
\bibitem [{\citenamefont {Newcomb}(1881)}]{10.2307/2369148}%
  \BibitemOpen
  \bibfield  {author} {\bibinfo {author} {\bibfnamefont {S.}~\bibnamefont
  {Newcomb}},\ }\href {http://www.jstor.org/stable/2369148} {\bibfield
  {journal} {\bibinfo  {journal} {American Journal of Mathematics}\ }\textbf
  {\bibinfo {volume} {4}},\ \bibinfo {pages} {39} (\bibinfo {year}
  {1881})}\BibitemShut {NoStop}%
\bibitem [{\citenamefont {Benford}(1938)}]{10.2307/984802}%
  \BibitemOpen
  \bibfield  {author} {\bibinfo {author} {\bibfnamefont {F.}~\bibnamefont
  {Benford}},\ }\href {http://www.jstor.org/stable/984802} {\bibfield
  {journal} {\bibinfo  {journal} {Proceedings of the American Philosophical
  Society}\ }\textbf {\bibinfo {volume} {78}},\ \bibinfo {pages} {551}
  (\bibinfo {year} {1938})}\BibitemShut {NoStop}%
\bibitem [{\citenamefont {Cong}\ \emph {et~al.}(2019)\citenamefont {Cong},
  \citenamefont {Li},\ and\ \citenamefont {Ma}}]{CONG20191836}%
  \BibitemOpen
  \bibfield  {author} {\bibinfo {author} {\bibfnamefont {M.}~\bibnamefont
  {Cong}}, \bibinfo {author} {\bibfnamefont {C.}~\bibnamefont {Li}}, \ and\
  \bibinfo {author} {\bibfnamefont {B.-Q.}\ \bibnamefont {Ma}},\ }\href
  {\doibase https://doi.org/10.1016/j.physleta.2019.03.017} {\bibfield
  {journal} {\bibinfo  {journal} {Physics Letters A}\ }\textbf {\bibinfo
  {volume} {383}},\ \bibinfo {pages} {1836 } (\bibinfo {year}
  {2019})}\BibitemShut {NoStop}%
\end{thebibliography}%


\appendix

\section{On the Assumptions of Stationarity, Poissonianity, and Independence}
\label{sec:assumptions}

The pointy statistic is based on the assumptions of stationarity and Poissonianity.
In practice, both of these may break down.
Although the pointy statistic remains useful in such cases, we briefly discuss a few diagnostics that may help address which assumption is breaking down.

Specifically, the assumptions of stationarity and Poissonianity are encoded in the distribution of time between consecutive events (Eqn.~\ref{eq:Delta t}).
By directly examining the observed distribution of the time between events, and modeling how this distribution changes over time, one could empirically remove the need for these assumptions.
However, data sets may be sparse enough that some level of modeling will be needed, whether in the form of the distribution or how rapidly it can vary over time.
A stationary Poisson process is a reasonable approximation for the data we consider.

\begin{figure}
    \begin{center}
        \includegraphics[width=1.0\columnwidth, clip=True, trim=0cm 1.20cm 0cm 0.00cm]{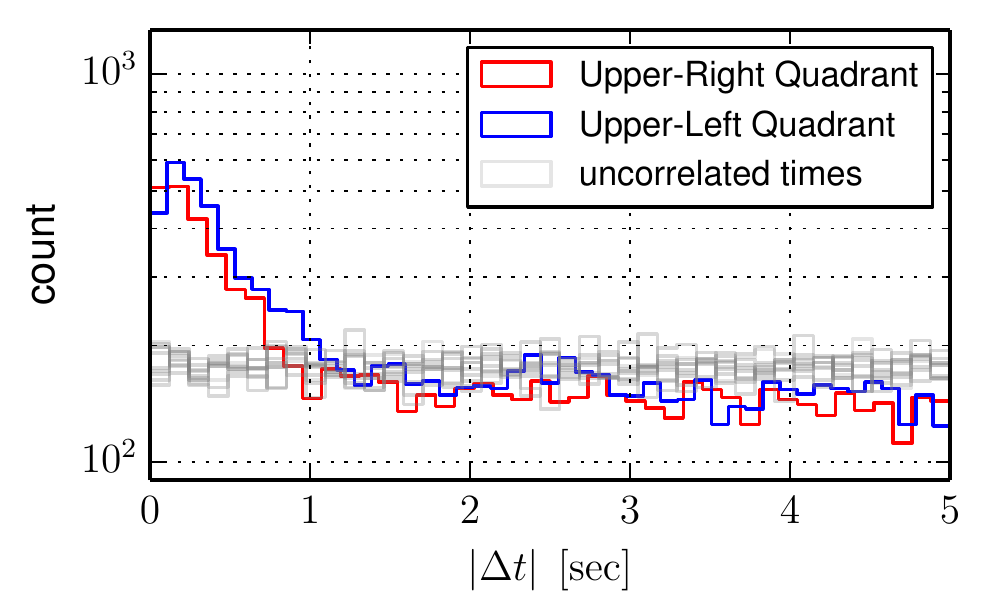}
        \includegraphics[width=1.0\columnwidth, clip=True, trim=0cm 0.05cm 0cm 0.10cm]{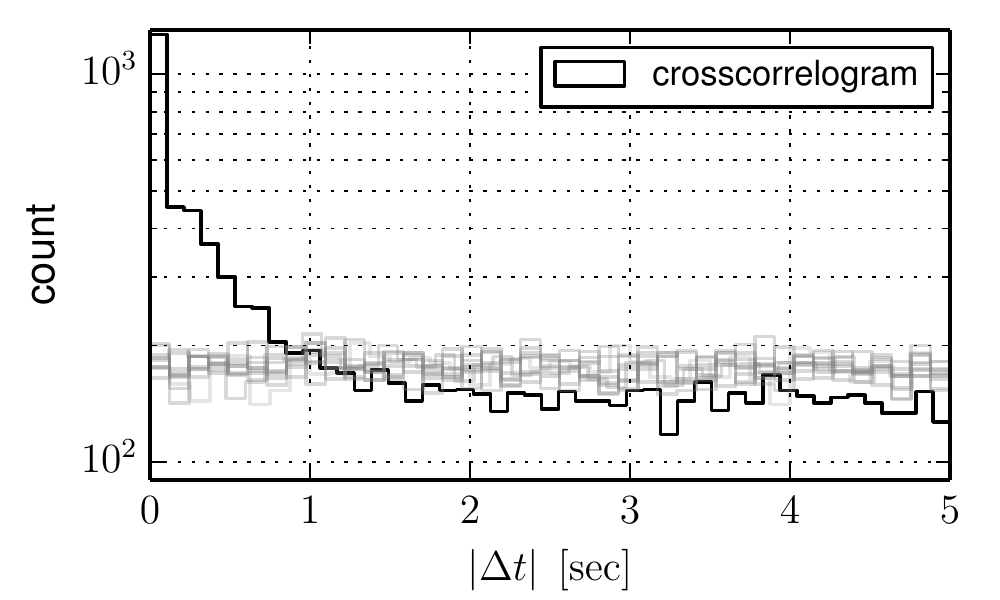}
    \end{center}
    \caption{
        (\emph{top}) Separate autocorrelograms for events from two channels measuring the current through two nearby electrostatic actuators in the power-recycling mirror suspension system of LIGO Livingston Observatory between 8-128$\,\mathrm{Hz}$: (\emph{red}) upper-right and (\emph{blue}) -left quadrants of the second stage: \result{L1:SUS-PR2\_M2\_FASTIMON\_\{U,L\}R\_OUT\_DQ} along with (\emph{grey}) distributions from the same number of randomly distributed events.
        (\emph{bottom}) The crosscorrelogram between the two channels.
        Although there is an excess of nearby events compared to the randomly distributed times (the channels produce correlated clusters of events), the channels are well modeled as independent Poisson processes on timescales $\gtrsim \mathcal{O}(1\,\mathrm{sec})$.
    }
    \label{fig:correlograms}
\end{figure}

Correlograms, or histograms of the time between events, can be useful diagnostic tools.
Correlograms either compare all events within a single set (autocorrelogram) or events from two separate sets (crosscorrelogram).
Fig.~\ref{fig:correlograms} demonstrates examples of each.
Comparing these with our expectations from Poissonianity and independence provide a valuable sanity check.
For a Poisson process all the pair-wise time differences between events, i.e., the autocorrelogram, are expected to form a uniform distribution.
This remains the case for the crosscorrelogram when constructed from two \emph{independent and uncorrelated} Poisson processes.
For example, channels may be Poisson distributed but correlated, in which case the autocorrelograms would agree with expectations but the crosscorrelogram would not.
Similarly, non-Poissonianity in a single channel would appear in that channel's autocorrelogram, but the crosscorrelogram could still appear as expected if the channels are independent.

Furthermore, we sometimes assume statistical independence between different channels.
This allows us to analyze each channel independently and then stack \pvalues~by multiplying results obtained from different channels to create a na{\"i}ve Bayes detection statistic: $P_\mathrm{joint}$.
This assumption may hold quite well in many cases, such as different detector subsystems that are isolated and do not interact (e.g., sensors in different end stations within kilometer-scale GW interferometers).
Nonetheless, this assumption can often break down in practice.
For example, many auxiliary channels within GW detectors record data from nearly identical sensors (e.g., photodiodes are often divided in half or in quadrants).
Such channels will almost certainly witness correlated signals and therefore produce correlated sequences of events.
Similarly, multiple sensors often observe the same suspension system within ground-based GW interferometers, meaning that a single jolt of excess ground-motion could appear in all of them.
As an extreme example, actuation signals within control loops are often explicitly constructed as linear combinations of the sensor data.
Ref.~\cite{Essick2020} discusses the fully-connected probabilistic graphical model that describes the complicated situation within real detectors.

\begin{figure}
    \includegraphics[width=1.0\columnwidth]{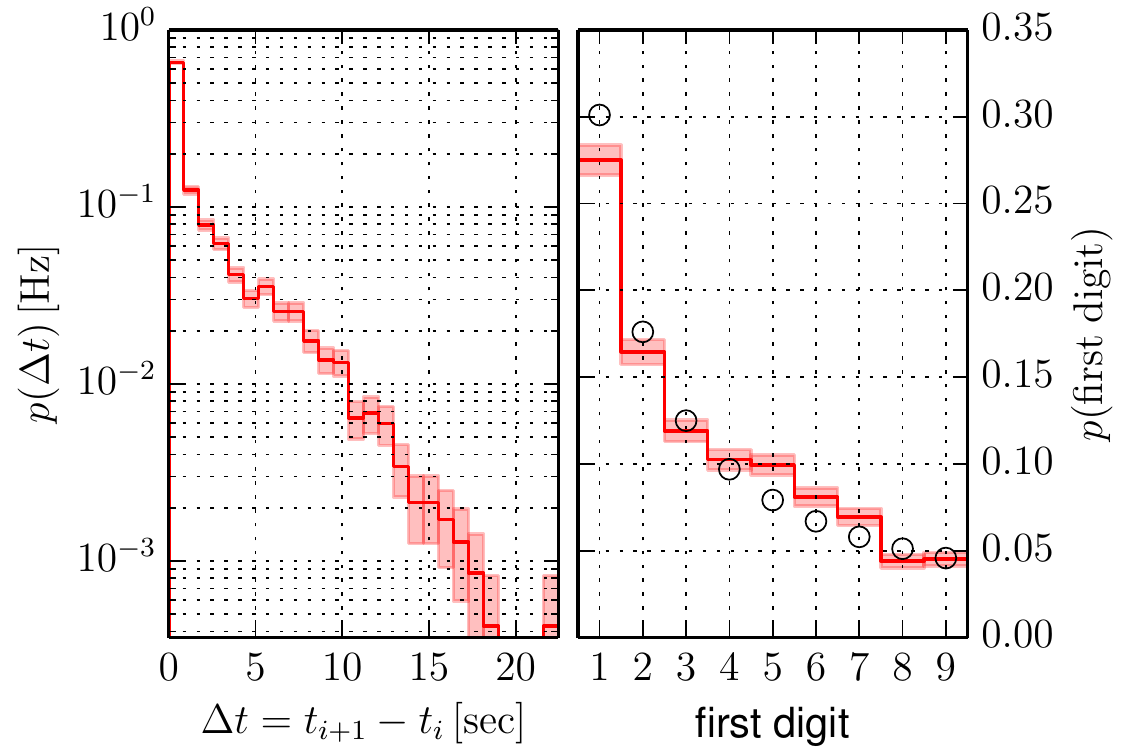}
    \caption{
        (\emph{left}) Observed distribution of the time between consecutive events in the upper-right quadrant of the power-recycling mirror suspension system shown in Fig.~\ref{fig:correlograms}.
        Shaded regions approximate 1-$\sigma$ counting uncertainty.
        We note that the data is roughly exponentially distributed, as expected, but with an excess at $\Delta t \lesssim 1\, \mathrm{sec}$, in agreement with Fig.~\ref{fig:correlograms}.
        (\emph{right}) The distribution of the first digit (\emph{red}) and the analytic predictions from Benford's law (\emph{black circles}).
    }
    \label{fig:benford}
\end{figure}

Benford's law~\cite{10.2307/2369148, 10.2307/984802}, or the relative frequency of the first digit ($d$) of each number in a large set,\footnote{For example, $d=1$ for $13$ and $d=8$ for $831$.} provides an additional test of whether a channel's behavior is expected.
In a base-10 number system, Benford's law predicts that the relative frequency of the first digit in a number drawn is given by
\begin{equation}
    P(d) = \log_{10}\left( 1 + \frac{1}{d}\right)
\end{equation}
Indeed, the time-between-events in a Poisson distribution is exponential, and should closely follow Benford's law~\cite{CONG20191836}.
Fig.~\ref{fig:benford} demonstrates this for one of the channels shown in Fig.~\ref{fig:correlograms}.
As with the correlogram, we see that there are deviations from the expected distribution, but the data is reasonably approximated by the analytic prediction nonetheless.

While the pointy statistic's assumptions may break down, Sec.~\ref{sec:applications} shows how to directly measure how $P_\mathrm{min}$ and/or $P_\mathrm{joint}$ are distributed in the presence of correlated channels within real data sets.
This means we can still make rigorous, statistically precise statements regardless of whether or not the assumptions that motivate the functional form of the pointy statistic and our na{\"i}ve Bayes techniques hold exactly.

We would be remiss if we did not discuss at least one additional subtlety.
The pointy statistic's assumption of Poissonianity implies events within a channel are completely uncorrelated.
If this is violated by, for example, a preference for events to occur regularly on integer second boundaries, then selecting times-of-interest that are also regularly spaced with the same periodicity may introduce spuriously small $P_\mathrm{min}$.
That is to say, the pointy statistic detects that events occur too close to the selected times to be due to random chance, but this is because we have accidentally aligned our selected times with preexisting correlations within that channel.
Indeed, correlation does not imply causation.
Because we often wish to infer a common cause based on coincidence null tests, care should be taken to randomly distribute the times of interest as much as possible to avoid accidentally coinciding with existing periodic behavior within series of events that are not Poisson distributed.
This is particularly relevant for hardware injections to determine auxiliary channel safety within GW interferometers, where analysts have complete control over the parameters of each injection.

\end{document}